# Gate-tuned ambipolar superconductivity with strong pairing interaction in intrinsic gapped monolayer 1T'-MoTe$_2$

Fangdong Tang[1], Peipei Wang[2], Yuan Gan[2], Jian lyu[2], Qixing Wang[1], Xinrun Mi[3], Mingquan He[3], Liyuan Zhang[2], Jurgen H. Smet[1*]

[1] Max Planck Institute for Solid State Research, Stuttgart 70569, Germany.

[2] Department of Physics and Shenzhen Institute for Quantum Science and Engineering, Southern University of Science and Technology, Shenzhen 518055, China.

[3] Low Temperature Physics Laboratory, College of Physics & Center of Quantum Materials and Devices, Chongqing University, Chongqing 401331, China.

* Corresponding author. Email: j.smet@fkf.mpg.de

**Gate tunable two-dimensional (2D) superconductors offer significant advantages when studying superconducting phase transitions. Here, we address superconductivity in exfoliated 1T'-MoTe$_2$ monolayers with an intrinsic band gap of ~7.3 meV using electrostatic doping. Despite large differences in the dispersion of the conduction and the valence bands, superconductivity can be achieved easily for both electrons and holes. The onset of superconductivity occurs near 7-8K for both charge carrier types. This temperature is much higher than in bulk samples. Also the in-plane upper critical field is strongly enhanced and exceeds the BCS Pauli limit in both cases. Gap information is extracted using point-contact spectroscopy. The gap ratio exceeds multiple times the value expected for BCS weak-coupling. All these observations suggest a strong enhancement of the pairing interaction.**

## INTRODUCTION

Gate tunable 2D superconductors allow a systematic study of the superconducting phase transition as a function of the carrier density without the need of producing multiple samples with different compositions using complex crystal growth. While in the majority of cases either the very potent ionic liquid gating technique (*1*, *2*) or the proximity effect (*3*, *4*) is required to induce superconductivity, there are also a few materials that exhibit intrinsic superconductivity at low densities that can be more conveniently reached with conventional field effect gating using a solid dielectric gate. Reported examples include oxide heterostructure interfaces (*5*, *6*), twisted bilayer graphene (*7*, *8*) and 1T'-WTe$_2$ monolayers (*9*, *10*). The last are a member of the family of transition metal dicalchogenides (TMDs MX$_2$, with M=Mo or W and X=S, Se, or Te) that come in different phases. Recently, the 1T'-phase and the T$_d$-phase have been intensively investigated. They are predicted to be type-II Weyl semi-metals in the bulk form (*11*, *12*) and large-gap 2D topological insulators (TI) in the monolayer form (*13*). The band gap opening in the monolayer was predicted as the combined result of band inversion and spin-orbit coupling (SOC) (*13–15*). The 1T'- and T$_d$-phases share the same in-plane crystal structure, but possess a different out-of-plane stacking sequence. The 1T'-phase has a monoclinic shape, while the T$_d$-phase is orthorhombic (*16*, *17*). In the remainder, we focus on the 1T'-MoTe$_2$ phase. Different to the bulk system with a superconducting transition temperature of about 0.1 K (*17*), the critical temperature is significantly enhanced in few-layer samples implying an increasing strength of the pairing interaction with reduced thickness (*18–20*). This behavior with thickness is extremely rare in 2D superconductors. While several experimental studies have suggested insulating behavior in monolayer 1T'-MoTe$_2$, from which the authors concluded the existence of a gap in the bulk (*16*, *21*), other works have claimed semi-metallic behavior and the absence of a gap in the bulk instead (*14*, *15*, *20*). Disorder and sample degradation during sample processing, handling and mounting (*21*, *22*) may be responsible for these contradictory results and hence continued efforts to unveil the intrinsic physics of the monolayer are highly desirable.

In this work, we provide unambiguous evidence that the monolayer 1T'-MoTe$_2$ is an insulator. We also demonstrate ambipolar superconductivity for electrons in the Q-valleys and holes near the Γ-point in the valence band. The onset of superconductivity increases all the way up to ~7.5 K and the in-plane critical field $B_{c2,\parallel}$ for both the Q- and Γ-Fermi pockets exceeds the BCS Pauli limit (*23*, *24*). Since the Fermi surface topology of the valence and conduction band are quite distinct, the enhanced stability of the superconducting properties is likely not related to the band structure, but rather the interaction among the mating charge carriers forming a Cooper pair. This is confirmed in point-contact measurements.

## RESULTS

**Band gap opening in monolayer samples**

Our measurements were performed on 1T'-MoTe$_2$ devices fabricated from high-quality single crystals synthesized with the flux method using NaCl (*16*, *18*). Unless explicitly mentioned, in the main text only monolayer devices (D1-D5) are discussed. Four of these (D2-D5) have both a back-gate (the doped Si substrate) as well as a top-gate. The latter consists of a graphitic layer on top of the encapsulating hBN layer. A schematic of the layer sequence and geometry is illustrated in Fig. 1A. Several optical images of completed devices are shown in Fig. S1 in the Supplementary Material (SM). The field effect induced density can be calculated from $n = (V_{tg}c_{tg} + V_{bg}c_{bg})/e$, where $c_{tg}$ and $c_{bg}$, are the areal capacitances for the top and bottom gate, respectively, $e$ is the electron charge, $V_{tg}$ the top-gate voltage and $V_{bg}$ the back-gate voltage (Fig. S3, Section S3 of the SM).

Fig. 1B schematically depicts the band structures of monolayer and multilayer 1T'-MoTe$_2$ (*13–16*, *19–21*). The band inversion arises from the period doubling of the Mo chain in the 1T' structure. It lowers the Mo d-orbital below the Te p-orbital near Γ. The presence of SOC opens a topological band gap in the monolayer (*13–15*). The on-site Coulomb interaction is reported to influence significantly the electronic band structure (*14*, *25*). This may be a reason for the observation of different electronic phases reported in various experiments (*14*, *15*, *20*). The temperature dependence of the sheet resistivity $\rho_s$ for samples of different layer thicknesses is summarized in Fig. 1C (see Section S2 of SM for more details). The sheet resistivity increases significantly when the layer thickness decreases from four layers to one layer. In the insert to Fig. 1C, the temperature at which the resistance starts to drop has been marked for the samples with different layer thickness. This signals the onset of superconductivity. The temperature at which this occurs systematically increases with decreasing layer thickness. In the remainder of the manuscript, we will primarily use the temperature where the resistance has dropped to 50% of the normal state resistance as the transition temperature. It will be referred to as $T_{c,0}$ in the absence and as $T_c$ in the presence of an external applied magnetic field. For thicknesses above two layers, the samples display metallic behavior and the critical temperature $T_{c,0}$ remains below 3 K. For monolayer devices, the temperature dependent behavior of the resistance depends on the charge carrier density. At high carrier densities, samples behave metallic across the entire temperature range and become superconducting below 7.5 K. For low density, samples undergo a metal-insulator transition. This heralds the opening of a band gap. The resistance drops sharply near 7.5 K signaling the superconducting transition. Such a low density case is illustrated in Fig. 1C. When the charge carrier density is reduced further, the band gap fully develops and the sample remains insulating down to the lowest temperature (see Fig. 2A and Fig. S4C in SM of device D4). The gap size is estimated from an Arrhenius fit in Section 6 of the SM. This analysis yields a gap value of ~7.3 meV. At temperatures above 100 K (~8.7 meV), thermally activated charge carriers dominate the transport. In this temperature regime the behavior is metallic because of the reduction of phonon scattering with decreasing temperature. The metal-insulator transition (MIT) happens as the bulk gap gradually opens below 100 K. Unfortunately, there is no evidence for topologically non-trivial states that one may anticipate due to band inversion (*13*). Exemplary data recorded at large carrier densities on monolayer devices that remain metallic are discussed in Section 4 of the SM. The shift to higher temperature of the onset of superconductivity, marked by arrows in the inset Fig. 1C, is consistent with previously reported results on few layer and monolayer devices (*18*, *20*). Since the theoretical calculations on monolayer 1T'-MoTe$_2$ predict single band behavior without any signature of a density-wave induced modulation of the density of states, this temperature enhancement is most likely a consequence of increased pairing interactions (*20*, *26*, *27*). This will be elaborated upon near the end of this letter.

**Gate-tuned ambipolar superconductivity**

Fig. 1D displays a 2D color map of the gate-dependent resistance measured on monolayer device D3 at a temperature of 1.8 K and at zero magnetic field. As the charge carrier density is tuned by sweeping the top and bottom gate voltages, the sample undergoes a transition from a p-type (around the Γ-point) superconducting state to an insulating state with a gap inside the bulk and finally a n-type superconducting state (around the Q-point). The superconducting behavior is observed for electron and hole densities above $7 \times 10^{12}$ cm$^{-2}$. This contrasts with 1T'-WTe$_2$ monolayers where superconductivity only appears for electron doping (*9*, *10*). A single line trace crossing the insulating state is shown in the inset of Fig. 1E. Data for two additional devices D2 and D4 can be found in Section S3 of the SM (Fig. S3). The low-temperature resistance of the insulating state reaches values as high as ~$10^5$ ohms for device D4 and ~$10^4$ ohms for other devices. Fig. 1E illustrates the temperature dependence of the resistance for three different carrier densities: a hole density of $-12.0 \times 10^{12}$ cm$^{-2}$, an electron density of $18.8 \times 10^{12}$ cm$^{-2}$ and an estimated net hole carrier density of $-0.2 \times 10^{12}$ cm$^{-2}$. The latter density is only a rough estimate. Hall measurements are problematic in this regime, since the sample is in an insulating state (*9*, *20*) and most likely both charge carrier types are simultaneously present across the sample area due to inhomogeneous disorder. For the trace with a hole density of $-12.0 \times 10^{12}$ cm$^{-2}$, the resistance starts to drop around ~7.5 K from its normal state value. It reaches 50% of the normal state resistance near ~6.2 K and has fallen to zero below ~5 K. For an electron density of $18.8 \times 10^{12}$ cm$^{-2}$, the onset of superconductivity is also around ~7.5 K, but the resistance goes down more slowly and reaches the 50% mark of the normal state resistance near ~4.5 K. Below ~2.4 K, the resistance is close to zero. For the trace with small net hole doping ($-0.2 \times 10^{12}$ cm$^{-2}$), the resistance develops a local maximum around ~7.5 K, but then continues to rise in a non-monotonous fashion at lower temperature. This apparent insulating behavior will be discussed in more detail later. Fig. 1F plots the differential resistance d$V$/d$I$ as a function of the dc bias current $I_{dc}$ for the three different carrier densities marked in this panel. For large electron or hole doping, the sample is in the superconducting state with zero differential resistance at low bias current. The sharp peak at low bias current for a low net electron doping of $1.1 \times 10^{12}$ cm$^{-2}$ is indicative of an insulating ground state instead.

**Electronic phase diagram**

Here we intent to compile a diagram of the different electronic states that emerge as the density is tuned. Fig. 2A illustrates the evolution of the temperature-dependent resistance for monolayer sample D4. For a low net hole doping of $-8 \times 10^{11}$ cm$^{-2}$ clear insulating behavior is observed. This behavior persists also for low net electron doping, but the resistance decreases with increasing electron density. Eventually the system behaves metallic and at sufficiently high electron density, superconducting behavior appears. It is very pronounced at the highest shown density of $2.21 \times 10^{13}$ cm$^{-2}$. The transition from the insulating state to the superconducting state with carrier density occurs continuously with no clear critical density. Additional data for other devices are shown in Section S4 of the SM. For the data recorded on device D4 in Fig. 2A, the resistance remains non-zero down to the lowest temperature and the highest accessible density, despite the clear superconducting transition. This is not uncommon for low-density 2D superconductors and is usually attributed to insufficient carrier density, filtering, cooling or also sample purity. Strictly speaking we are dealing with a "failed superconductor" usually referred to in the literature as an anomalous metal or a quantum metal that can be tuned by electrostatic gating, magnetic field or changes in the degree of disorder (*2*, *28*, *29*). These systems typically exhibit quantum critical behavior with magnetic field caused by quantum fluctuations in the inhomogeneous superconductor in which superconducting regions are rare (Section S11 of SM) (*29*, *30*). Some of the data traces in Fig. 2A have been replotted in Fig. 2B on a logarithmic resistance and 1/T scale. By presenting the data in this fashion, it is possible to identify the transition from thermally activated flux flow (linear-dependence marked with black lines) to the anomalous metal regime where the resistance saturates as the temperature is lowered (*2*, *28*).

Differential resistance data recorded in the dc bias current versus density parameter plane are also very helpful to identify transitions between different electronic phases. An example is shown in Fig. 2C. This color map of the differential resistance was acquired on monolayer device D3 at a temperature of 1.8 K. Additional data sets are available in Section S6 of the SM. For low net densities when $|n| \lesssim 7 \times 10^{12}$ cm$^{-2}$, a sharp peak is observed in the differential resistance at low bias current. It signals the insulating ground state. As the carrier density is increased beyond $|n| \gtrsim 7 \times 10^{12}$ cm$^{-2}$, the amplitude of this zero-bias peak gradually drops. When entering the superconducting regime, the differential resistance remains zero. The region of zero differential resistance becomes wider as the superconducting ground state gets better developed at larger carrier densities. For larger bias currents, additional differential resistance peaks are observed. The bias currents at which they appear are not constant but vary in a continuous but erratic manner on carrier density. This may be related to a rearrangement of the insulating and superconducting regions as the carrier density is tuned. The non-monotonous behavior of the *R* vs. *T* curves in Fig. 1E and Fig. S4 also are indicative of such a scenario. Fig. 2D illustrates the density dependence of the magnetoresistance recorded on device D4 in a perpendicular magnetic field at 1.7 K. Here, the insulating state and superconducting states can be distinguished better, because the resistance value changes more dramatically in the three different regimes. The out-of-plane critical field $B_{c2,\perp}$ at higher densities is on the order of ~2 T. This is shown in Fig. S11B and Table S1. This critical field is significantly larger compared to that of most low-density superconductors (~0.1 T). Fig. 2E plots the evolution of the gate-dependent resistance with temperature in the range from 1.8 K to 15 K at zero magnetic field for monolayer device D3. The sequence of phase transitions from a p-doped superconducting state to an insulating state and from the insulating state to an n-doped superconducting state are visible (see also Section S7 for more data). There is no clear critical density $n_c$ as observed in monolayer 1T'-WTe$_2$ (*9*). Instead, there is a transition region between $3\times 10^{12} <|n|< 7\times 10^{12}$ cm$^{-2}$ that forms the phase boundary between the insulating and superconducting state. In the insulating phase there is no specific evidence for the existence of topological nontrivial edge states within the bulk gap as predicted for such a system with band inversion and SOC (*13*). This would show up as a plateau in the resistance with a quantized value of $h/2e^2$ (*31*, *32*). In the insulating regime the resistance typically exhibits strong fluctuations and depending on the quality of the sample, resistance values as large as ~10$^5$ ohms are observed in device D4. Further details on the insulating state can be found in Section S7 and S8 of the SM. In the low density region, insulating and superconducting regions coexist. The latter are linked by Josephson-like coupling, which varies as the carrier density is tuned.

A diagram summarizing the different phases detected in device D3 is displayed in Fig. 2F. For other samples we refer to Section S4 of the SM for similar diagrams. In sample D3 the insulating state around zero average carrier density persists up to $|n|<3\times 10^{12}$ cm$^{-2}$ with superconducting states on either side for electron and hole doping. They convert into normal metallic states above the transition temperature. We note that the superconducting phases for electron and hole doping emerge differently out of the insulating phase. For hole carriers, the transition to the superconducting state is much sharper and clearer. The same is apparent from the behavior of the critical current, the critical field as well as transition temperature when tuning the carrier density (see Fig. 2C-F). This difference between hole and electron doping can be understood from the band structure of monolayer 1T'-MoTe$_2$. The curvature of the valence band at the Γ-point is much smaller than that of the conduction band at the Q symmetry points of the Brillouin zone. Hence, the effective mass and the density of states are significantly larger for hole doping. We note that the temperature at which we observe the onset of superconductivity varies little with the average density. Cooper pair formation occurs as long as the temperature is below $T_{c,onset}$ as superconducting puddles even exist at low average density due to the spatial density inhomogeneity.

**Evidence for strong pairing interaction**

In the following, we focus on the unconventional nature of the superconductivity in these 1T'-MoTe$_2$ monolayers. Fig. 3A illustrates the temperature dependent resistance recorded on device D2 for different perpendicular ($B_\perp$, left panel) and parallel magnetic fields ($B_\parallel$, right panel) and a fixed charge carrier density of $2.38 \times 10^{13}$ cm$^{-2}$. The in-plane and out-of-plane upper critical fields are shown as a function of the normalized temperature in Fig. 3B with blue circles ($B_{c2,\parallel}$) and black squares ($B_{c2,\perp}$). The temperature is normalized with $T_{c,0}$. The experimental data is fitted to the 2D Ginzburg-Landau (GL) model in order to extract the zero temperature Ginzburg-Landau in-plane coherence length $\xi_{GL}$ using the expression $B_{c2,\perp}(T) = \frac{\phi_0}{2\pi\xi_{GL}^2}(1 - T/T_{c,0})$. Here $\phi_0 = \frac{h}{2e}$ is the magnetic flux quantum for Cooper pairs (*33*). A least square fit to the data points in Fig. 3B yields $\xi_{GL}(0K) \sim 12.1$ nm. The BCS coherence length $\xi_0$ then follows from the relationship $\xi_0 = 1.35\xi_{GL}(0K)$ (*33, 34*) or $\xi_0 \sim 16.3$ nm. The mean free path of the charge carriers is estimated from the Drude model, $l_{mfp} = h/(e^2\rho_s\sqrt{g_s g_v \pi n_e})$, where $\rho_s$ is the sheet resistivity, $n_e$ is the net charge carrier density, and $g_s = g_v = 2$ for spin and valley degeneracy in the case of electrons (*10, 20*). We obtain a mean free path $l_{mfp} \sim 7.8$ nm. Since $\xi_0 > l_{mfp}$, the superconductor is in the dirty limit. For the in-plane critical field, a phenomenological square-root dependence of $B_{c2,\parallel}(T) \propto (1 - T/T_{c,0})^{1/2}$ is used to fit the in-plane field data near $T_{c,0}$ (*33, 35*). Such a fit yields $B_{c2,\parallel}(0K) \sim 19 \pm 1$ T. This value is about 2 times larger than the BCS Pauli limit $B_p^{BCS}(0K) = 1.85 T_{c,0} \sim 8.7$ T that has been marked with a horizontal black dashed line. This is consistent with previous reports (*18–20*). Similar behavior is observed for holes. In Fig. 3D, data are shown for a hole density of $-1.10 \times 10^{13}$ cm$^{-2}$. Here too, $B_{c2,\perp}(T)$ exhibits a linear temperature dependence as well as an enhanced $B_{c2,\parallel}(0K)$ compared to the BCS Pauli limit. The angular dependence of the upper critical field ($B_{c2}(\theta)$) at a temperature of 1.65 K is plotted in Fig. 3C by rotating the sample with respect to the direction of the external applied magnetic field. The critical field is defined here as the field at which the resistance has risen to 35% of the normal state resistance. The data points were fitted to the 2D Tinkham model (black line) as well as the 3D anisotropic Ginzburg-Landau model (blue line) (*19, 33*). The former model achieves better agreement with the data. Also the divergent behavior of the $B_{c2,\parallel}/B_{c2,\perp}$ ratio when approaching $T_{c,0}$ plotted in the insert to Fig. 3B corroborates the 2D nature of the superconductivity (*36, 37*). The key density dependent superconducting parameters such as $T_{c,0}$, $B_{c2,\perp}$, $\xi_0$ and $l_{mfp}$, are summarized in Section S9 of the SM (Fig. S11).

The origin of the enhanced in-plane upper critical field $B_{c2,\parallel}$ requires further discussion. In traditional type-II superconductors, the critical field $B_{c2,\parallel}$ is either governed by the paramagnetic effect or the orbital effect. Pair breaking can be induced by the orbital effect when a flux quantum penetrates the cross-sectional area determined by the sample thickness $d$ and the BCS coherence length, which characterizes the size of a Cooper pair. This occurs at a field $B_{c2,\parallel}^{orb}(0K) = \frac{\phi_0}{2\pi\xi_0 d}$ (33). Since a monolayer of 1T'-MoTe2 has an atomic thickness of only $d \sim 0.35$ nm, $B_{c2,\parallel}^{orb}(0K)$ is estimated to be as high as $\sim 58\pm 8$ T for an electron density of $2.38 \times 10^{13}$ cm$^{-2}$. An alternative way to estimate $B_{c2,\parallel}^{orb}(0K)$ is using the slope of $B_{c2,\parallel}(T)$ near $T_{c,0} \sim 4.7$ K. The data in Fig. 3B yields $\left|\frac{dB_{c2,\parallel}}{dT}\right|_{T_{c,0}} \sim 20.7$ T/K and then $B_{c2,\parallel}^{orb}(0K) = 0.69 \cdot \left|\frac{dB_{c2,\parallel}}{dT}\right|_{T_{c,0}} \cdot T_{c,0} \sim 67$ T. This field is far away from the experimentally determined in-plane critical field and, hence, it is not the orbital effect that causes pair breaking in our samples in the presence of an in-plane magnetic field. This is anticipated for 2D superconductors.

The paramagnetic effect originates from the preferential orientation of the spin in the presence of the applied field expressed by the Zeeman energy. This energy difference between electrons of opposite spin forming a Cooper pair can only be accommodated, if it is smaller than the superconducting gap or binding energy. This criterion yields the Pauli limit for the applied magnetic field: $B_p(0K) = \frac{\Delta(0K)}{\sqrt{g}\mu_B}$ (23, 24, 38). Here, $\Delta(0K)$ is the zero-temperature superconducting gap, $g$ is the Lande $g$ factor and $\mu_B$ is the Bohr magneton. Since $g = 2$, for BCS superconductivity, $\Delta(0K) = \Delta_{BCS}(0K) = 1.76 k_B T_{c,0}$ and the BCS Pauli limit becomes $B_p^{BCS}(0K) = 1.85 T_{c,0}$. An enhancement of the critical field above this limit either implies a larger $\Delta(0K)$ due to stronger pair interactions or an effective reduction of the Zeeman energy. The latter can be caused by strong spin orbit scattering or by a strong perpendicular spin orbit field, if this field dominates the spin splitting rather than the real Zeeman field. The latter is for instance the case in systems with Ising pairing (35, 38–41).

A fit the of the in-plane upper critical field data using the Werthamer-Helfand-Hohenberg model that considers the spin-paramagnetic effect has been added to Fig. 3B. The model fits the data under the assumption that spin-orbit scattering is very weak: $\lambda_{so} = 0$ (38, 39, 42). The fit procedure yields a value of 5.3 for the Maki parameter $\alpha$. We take $\alpha = \sqrt{2} B_{c2}^{orb}(0K)/B_p(0K)$ and obtain $B_p(0K) \sim 17.3$ T. This should be compared with the BCS Pauli limit $B_p^{BCS}(0K) \sim 8.7$ T and hence we observe an enhancement of the Pauli limit by a factor of $\sim 2.0$. A similar analysis has been performed in Fig. 3D for a hole density of $-1.10 \times$ $10^{13}$ cm$^{-2}$. In this case, the enhancement amounts to ~ 2.2. In order to corroborate that this enhancement is not the result of strong spin orbit scattering, but rather a strengthened pairing interaction in this 2D superconductor, we attempt to perform point-contact measurements in order to extract information about the superconducting gap as a measure of the pairing strength. Fig. 4A schematically shows the setup used for these experiments. Here a dc bias current $I_{dc}$ is applied across the contact in parallel with a small ac bias current excitation, $I_{ac}$, with a frequency of 17.77 Hz and a root-mean-square amplitude between 1 and 5 nA depending on the contact resistance. The point contact is a van der Waals-type contact formed when placing the 1T'-MoTe$_2$ monolayer on top of one of the Au electrodes in the glovebox. The interface is clean and no additional measures are taken such as for instance the application of pressure (*43–45*) (see Section S12 of SM for more details). Fig. 4B displays several traces of the differential conductance $G_{contact} = I_{ac}/V_{ac}$ as a function of the dc voltage $V_{dc}$ recorded on the Au/1T'-MoTe$_2$ junction #1 of monolayer device D5 for an electron density of $1.38 \times 10^{13}$ cm$^{-2}$. We will refer to these junctions as contacts in the remainder. Temperature serves as an additional parameter. Fig. 4C replots some of the data at lower temperatures by normalizing to the 8K data. Similar normalized data are also shown for contact #2 in Fig. 4E. Contact #2 was selected as a second example, because the behavior is distinct from contact #1. For the latter the Au/1T'-MoTe$_2$ junction is in the insulating regime, whereas the second contact is in the metallic regime.

We employ the extended single-band isotropic *s*-wave Blonder–Tinkham–Klapwijk (BTK) model to fit the point contact spectroscopy data (*43–48*) to obtain the temperature-dependent gap amplitudes $\Delta(T)$. They are shown in Fig. 4D and 4F for contact #1 and contact #2. Included in these graphs is also the BCS model for the superconducting gap. The deviation of $\Delta(T)$ from the BCS relation in Fig. 4D may be caused by the spatial inhomogeneity of the superconductor (*49–51*). $\Delta(0K)$ is estimated to be $2.8 \pm 0.2$ meV for contact #1 and $2.4 \pm 0.3$ meV for contact #2. This allows us to determine a gap ratio $\Delta(0K)/k_B T_{c,\text{onset}}$ of ~4.3 and ~3.7, respectively. These values are significantly larger compared to the gap ratio of 1.76 for BCS weak-coupling. The BCS ratio for weak coupling is exceeded by a factor of 2.1 - 2.5. Hence, we conclude strong pairing interactions in 1T'-MoTe$_2$ monolayers (*20, 47*). These enhancement factors match well the observed enhancement of the in-plane upper critical field compared to the Pauli limit by a factor of 2.0-2.2. This makes a convincing case that the enhancement of $B_{c2,\parallel}(0K)$ can be attributed to the gap ratio enhancement. For the sake of completeness we remind that the superconductor is in the dirty limit and it can therefore not be fully excluded that also spin-orbit scattering plays some role. We believe however that spin orbit scattering should only have a secondary effect on the enhancement of $B_{c2,\parallel}(0K)$ in this work. In Section 10 of the SM, we address spin orbit scattering in more detail in the context of the KLB theory (*10*, *40*).

**DISCUSSION**

Finally, we compare important superconductivity parameters measured on 1T'-MoTe$_2$ monolayers with other 2D superconductors. We refer to Table I of the supplementary material. Superconductivity in 1T'-MoTe$_2$ monolayers occurs at a density as low as ~$5\times10^{12}$ cm$^{-2}$, whereas in transition metal dichalcogenides exhibiting Ising superconductivity the required density is typically much larger and on the order of ~$10^{14}$ cm$^{-2}$ (*35*, *41*, *52*, *53*). Although the density required to induce superconductivity in a 1T'-MoTe$_2$ monolayer is comparable to that in monolayer 1T'-WTe$_2$ (*9*, *10*) and twisted bilayer graphene (*7*, *8*), the onset of superconductivity occurs at a significantly larger temperature, $T_{c,\mathrm{onset}}$ ~7.5 K. The same holds for the upper critical field $B_{c2}$. It is also much larger than in these other two systems. The large enhancement of $B_{c2,\parallel}$, as observed here, has also been reported for other 2D superconductors that exhibit either type I or type II Ising pairing. Type I Ising pairing occurs in non-centrosymmetric superconductors with out-of-plane mirror symmetry such as for instance 2H-NbSe$_2$ (*35*), 2H-TaS$_2$ (*26*) and 2H-MoS$_2$ (*1*, *41*). A large perpendicular spin orbit field pins the spins in the out-of-plane direction. The external applied field must compete with this spin orbit field in order to reorient spins in the in-plane direction. This results in a significant increase of $B_{c2,\parallel}$. There have also been reports of a strong $B_{c2,\parallel}$ enhancement due to type II Ising pairing in materials such as few-layer stanene (*52*) as well as PbTe$_2$ (*53*), that are centrosymmetric but possess an unusual band structure near the Γ-point. Mating carriers reside in bands with different orbital indices and the bands are split due to spin orbit locking despite inversion symmetry. The symmetry and band structure properties of a 1T'-MoTe$_2$ monolayer neither fit that of Ising type I nor Ising type II superconductors. There is no out-of-plane mirror symmetry and inversion symmetry is present. Superconductivity appears both near the Γ- and Q-symmetry points of the Brillouin zone without bands of different orbital indices. The superconductivity in monolayer 1T'-MoTe$_2$ most likely has nothing to do with special symmetries or band dispersions. Instead, point contact measurements suggest that a strong pairing interaction is responsible for the enhanced robustness of the superconductivity. Further theoretical analysis and more advanced experimental techniques are needed to shed light on the origin of the strong pairing interaction.

**MATERIALS AND METHODS**

**Device fabrication**

High-quality 1T'-MoTe$_2$ single crystals were synthesized with the flux method using NaCl (*16, 18*). The devices were fabricated by placing mechanically exfoliated monolayers or few-layer films on contacts pre-patterned on top of a heavily doped Si substrate that is covered with a 300 nm thick dry thermal SiO$_2$. The pre-patterned contacts offer the advantage that it is possible to avoid exposure of the sample to air, solvents as well as elevated temperatures during sample fabrication, packaging, bonding, sample mounting and transfer of the sample to the vacuum environment of the cryostat system in which the measurements take place. While important to achieve the highest possible quality, this processing route does come with the disadvantage of a lower contact yield. A more detailed description of the device fabrication procedure can be found in Section 1 of the Supplementary Material.

**Electrical measurements**

The low-temperature magneto-transport measurements were performed in a dry cryostat system equipped with a superconducting magnet offering an axial field of up to 14 T and a variable temperature insert that allows temperature tuning between 1.6 K and 400 K. Low-frequency lock-in detection at 17.77 Hz was deployed for noise suppression to record longitudinal and transverse resistances. Measurement currents ranged between 1 nA and 10 µA depending on the two-point resistance of the sample. For both transport and point-contact measurements, the back-gate voltage, top-gate voltage and the dc bias current were applied with the help of a Keithley 2400 Source Measure Unit, a YOKOGAWA 7651 DC Source and an Agilent B2911A precision Source Measure Unit, respectively. The dc-voltage appearing as a result of the bias current was amplified by a voltage pre-amplifier and measured with an Agilent 34401A digital multimeter.

**Acknowledgments:**

The authors thank Yafei Ren, K. T. Law, Xu Du, Dong Zhao and Jianyu Xie for fruitful discussions. Technical assistance was provided by Shumao Xu, Sheng Yang, Johannes Geurs, Daniela Tabrea, Nils Gross, Steffen Wahl and Marion Hagel.

**Funding:**
DFG (SPP2244) and the Graphene Flagship program core 3 (J.H.S.)
National Key R&D Program of China (Grants No. 2022YFA1403700) (L.Z.)
Shenzhen Fundamental Research Fund for Distinguished Young Scholars (Grants No. RCJC2020071414435105) (L.Z.)
National Natural Science Foundation of China (11904040) (M.H.)
Chongqing Research Program of Basic Research and Frontier Technology, China (Grant No. cstc2020jcyj-msxmX0263) (M.H.)
Chinesisch-Deutsche Mobilitätsprogamm of Chinesisch-Deutsche Zentrum für Wissenschaftsförderung (Grant No. M-0496) (M. H.)

**Author contributions:**
F.T. and J.H.S. conceived the project. F.T. fabricated the devices and carried out the electrical measurements. P.W., Y.G. and J. L. grew the bulk 1T'-$MoTe_2$ crystals and performed preliminary measurements on bulk and multilayers. F.T. and Q.W. performed the Raman measurements. F.T., X.M. and M.H. analyzed the data and all authors discussed the results. F.T. and J.H.S. co-wrote the manuscript with the help from all the other authors.

**Competing interests:**
The authors declare that they have no competing interests.

**Data and materials availability:**
All data needed to evaluate the conclusions of the paper are included in the paper and/or the Supplementary Material. Additional data related to this paper may be obtained from the authors on reasonable request.

**Figures and Tables**

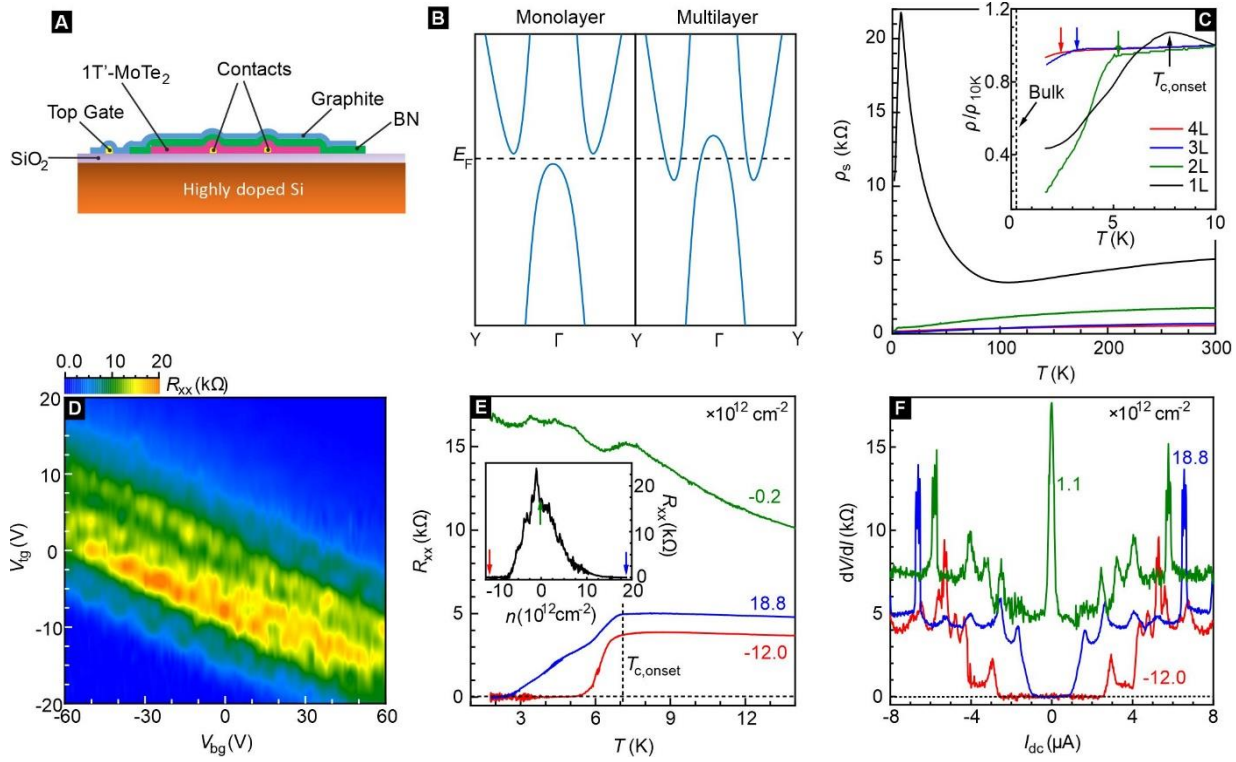

**Fig. 1. 1T'-MoTe$_2$ device geometry and basic transport characterization.** (**A**) Cross-sectional view of the device structure. (**B**) Schematic of the band structure for monolayer (left) and multilayer (right) 1T'-MoTe$_2$. (**C**) Temperature dependence of the sheet resistivity for samples with different layer thickness from 1 (1L) to 4 layers (4L). The inset shows a zoom of the sheet resistivity normalized to its value at 10 K. Arrows mark the temperature where the resistance starts to drop. This signals the onset of superconductivity. The temperature is referred to as $T_{c,\text{onset}}$. (**D**) Resistance measured at $T = 1.8$ K and $B = 0$ T as a function of the top and bottom gate voltages for monolayer device D3. (**E**) Temperature dependence of the resistance for three different charge carrier densities: $18.8 \times 10^{12}$ cm$^{-2}$ (electrons, blue), $-12.0 \times 10^{12}$ cm$^{-2}$ (holes, red) and $-0.2 \times 10^{12}$ cm$^{-2}$ (green, net estimated density). The insert shows the resistance as a function of carrier density for a fixed temperature $T = 1.8$ K. (**F**) Differential resistance (d$V$/d$I$) as a function of bias current for three different carrier densities.

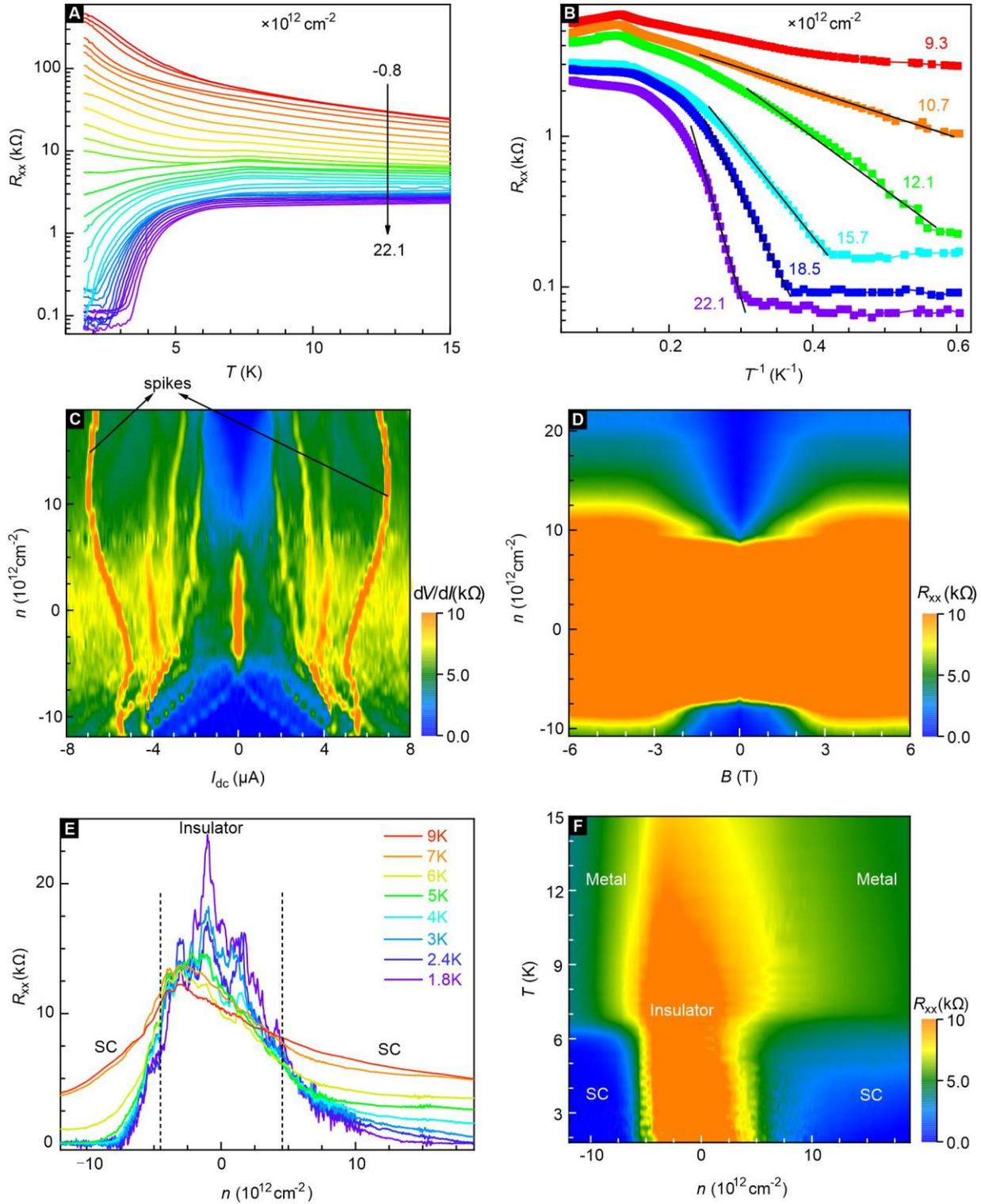

**Fig. 2. Longitudinal resistance and differential resistance as a function of the density, temperature and magnetic field.** (**A**) Temperature-dependent resistance at zero magnetic field with density as an additional discrete parameter. The transition from an insulating state to n-type superconductivity is clearly visible. Data were recorded on device D4. (**B**) Selected traces from panel (**A**) replotted on an inverse temperature scale. The black line is an Arrhenius fit describing the regime of thermally activated flux flow. Below this regime, we are dealing with the quantum metal state with finite resistance. Further details can be found in Section S11 of the SM. (**C**) Color map of the differential resistance $dV/dI$ in the parameter plane spanned by the dc bias current and the carrier density. Data were taken on device D3 at $T = 1.8$ K and $B = 0$ T. (**D**) Color map of the magnetoresistance for a perpendicular magnetic field and as a function of the carrier density. Data were taken on device D4 at $T = 1.7$ K. (**E**) Longitudinal resistance as a function of the density for different temperatures ranging from 1.8 K to 9 K.

The dashed black lines roughly mark the boundary between the insulating phase and the superconducting states. Data were taken on device D3 at $B = 0$ T. **(F)** Color map of the longitudinal resistance in the temperature vs. density parameter plane. The different states are marked. Single line traces from this data set are shown in panel **(E)**.

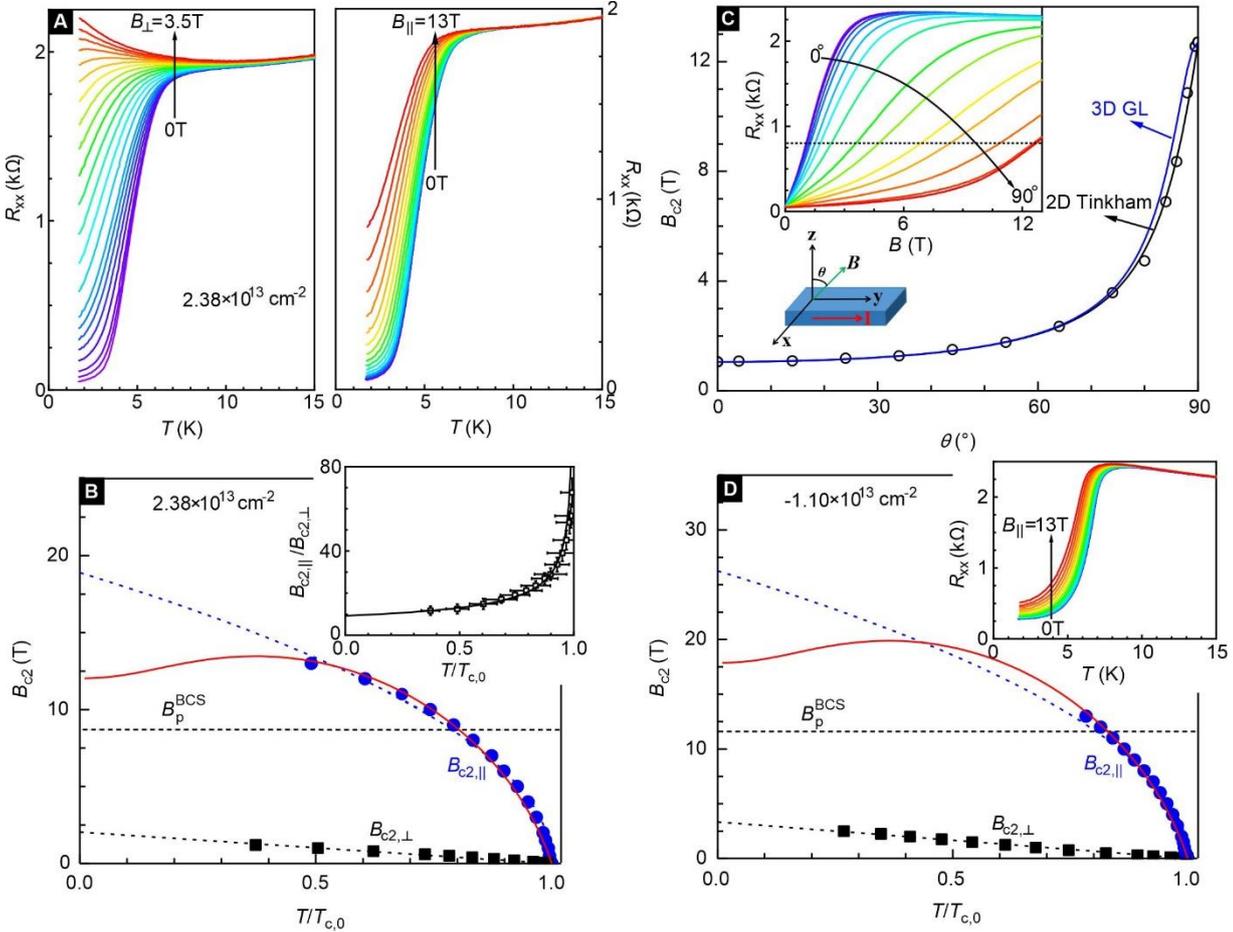

**Fig. 3. In-plane and out-of-plane magnetoresistance and critical field data.** All data were recorded on device D2. **(A)** Temperature dependent magnetoresistance for different perpendicular magnetic fields $B_\perp$ (left panel) and parallel magnetic fielda $B_\parallel$ (right panel). The data are recorded for an electron density of $2.38 \times 10^{13}$ cm$^{-2}$. **(B)** In-plane (black squares) and out-of-plane (blue circles) upper critical field as a function of temperature. The temperature is normalized to $T_{c,0}$. Dashed lines are fits of the upper critical fields using a Ginzburg-Landau phenomenological model. The solid red line is a fit of the in-plane upper critical to the Werthamer-Helfand-Hohenberg expression. The Pauli limit is shown by the black dotted line. The inset shows the ratio of the in-plane and out-of-plane upper critical field as a function of temperature. **(C)** Angular dependence of the upper critical field $B_{c2}$ for 1.65 K. For this panel, the critical field is defined as the field where the resistance has increased to 35% of the normal state resistance. The black and blue solid line are a fit with the 2D Tinkham formula $\left|\frac{B_{c2}(\theta)\cos\theta}{B_{c2,\perp}}\right| + \left(\frac{B_{c2}(\theta)\sin\theta}{B_{c2,\parallel}}\right)^2 = 1$ and the 3D anisotropic Ginzburg-Landau model $\left(\frac{B_{c2}(\theta)\cos\theta}{B_{c2,\perp}}\right)^2 + \left(\frac{B_{c2}(\theta)\sin\theta}{B_{c2,\parallel}}\right)^2 = 1$, respectively. The inset shows the original data of angular-dependent magnetoresistance at 1.65 K. **(D)** The same as **(B)** but for a hole density of $-1.10 \times 10^{13}$ cm$^{-2}$. The inset displays the temperature dependence of the resistance for different parallel magnetic fields.

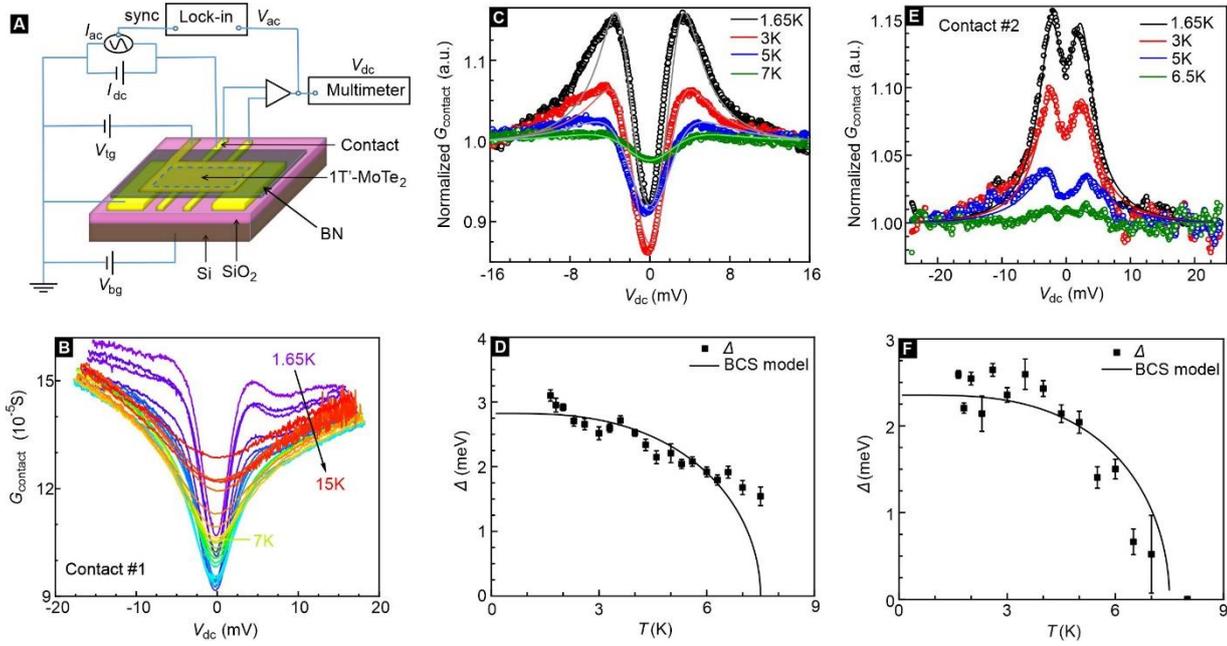

**Fig. 4. Point-contact spectroscopy in the superconducting regime.** All data were taken on device D5 for a density of $1.38 \times 10^{13}$ cm$^{-2}$ and for $B = 0$ T. (**A**) Schematic setup of the experimental arrangement used in the point-contact measurements. The applied ac bias current has frequency of 17.77 Hz. The amplitude is set between 1 to 5 nA depending on the contact properties. (**B**) Temperature evolution of the differential conductance $G_{contact}$ versus $V_{dc}$ recorded on contact #1. (**C**) Theoretical analysis of the temperature-dependent normalized $G_{contact}$ of contact #1 using the isotropic BTK model. (**D**) Temperature dependence of the gap amplitudes obtained from the BTK fit for contact #1. (**E and F**), Same as (**C** and **D**) but for contact #2.

## Supplementary Materials
Supplementary material for this article is available at http://xxxxxx

# Supplementary Materials for

## Gate-tuned ambipolar superconductivity with strong pairing interaction in intrinsic gapped monolayer 1T'-MoTe$_2$

Fangdong Tang, Peipei Wang, Yuan Gan, Jian lyu, Qixing Wang, Xinrun Mi,
Mingquan He, Liyuan Zhang, Jurgen H. Smet*

*Corresponding author. Email: j.smet@fkf.mpg.de

**This PDF file includes:**

# Section S1: Device fabrication and methods

Bulk single crystals of 1T'-MoTe$_2$ were synthesized using the flux method with NaCl (*16*, *18*). Monolayer and few layer films of 1T'-MoTe$_2$ were mechanically exfoliated from these bulk crystals using sticky tape inside a glovebox with a residual amount of water and oxygen of less than 0.1 ppm. Instead of transferring the flakes on the sticky tape directly onto the Si substrate, covered with the usual 300 nm thick dry thermal SiO$_2$ to analyze their thickness, a softer polydimethylsiloxane (PDMS) stamp was used as an intermediate to place the flakes on to such a Si substrate in order to obtain larger sized flakes. The flakes were typically not homogeneous in thickness, but were composed of the desirable monolayer or few layer region as well as an undesirable area with larger thickness. In order to remove the latter, a tear-and-release procedure using hBN on the PDMS stamp was used (*54*). If the hBN was large enough to cover the monolayer or few layer region of the 1T'-MoTe$_2$ flake and only touched a small portion of the substrate itself, the part of the 1T'-MoTe$_2$ flake in touch with the hBN was torn off from the thicker part and picked up. This approach required no heating or other steps that would be detrimental for the sample quality. The hBN/1T'-MoTe$_2$ heterostructure was then transferred onto a substrate with pre-patterned Cr(5nm)/Au(20nm) contacts. Subsequently a graphite film was put on top of the hBN. It covers the whole area of the 1T'-MoTe$_2$ film and also touches a separate contact. This sequence of fabrication steps is illustrated with optical images in Fig. S1A-F.

To avoid any exposure to ambient air, packaging on a suitable chip carrier and bonding of the device with indium wires also proceeded inside the glove box. The heavily doped Si substrate served as a back-gate. A schematic of the sample is illustrated in Fig. 1A of the main text and the optical images of three completed devices (D2-D4) are shown in Fig. S1G-I. With the help of a load lock arrangement, the sample was mounted into the sample rod and subsequently transferred to the cryostat without exposure to ambient air. The entire procedure avoided not only air exposure of the 1T'-MoTe$_2$ sample, but also any exposure to solvents or elevated temperatures, all of which would cause a degradation of the sample quality. An accurate determination of the MoTe$_2$ layer thickness was performed a posteriori by combining three pieces of information: the optical contrast produced by the flake during exfoliation and stacking, the onset of the superconducting transition and confocal Raman spectroscopy after completion of the magneto-transport measurements. The thickness determination works particularly well for monolayer and bilayer devices, as shown in Section 2, but becomes less reliable for thicker layers.

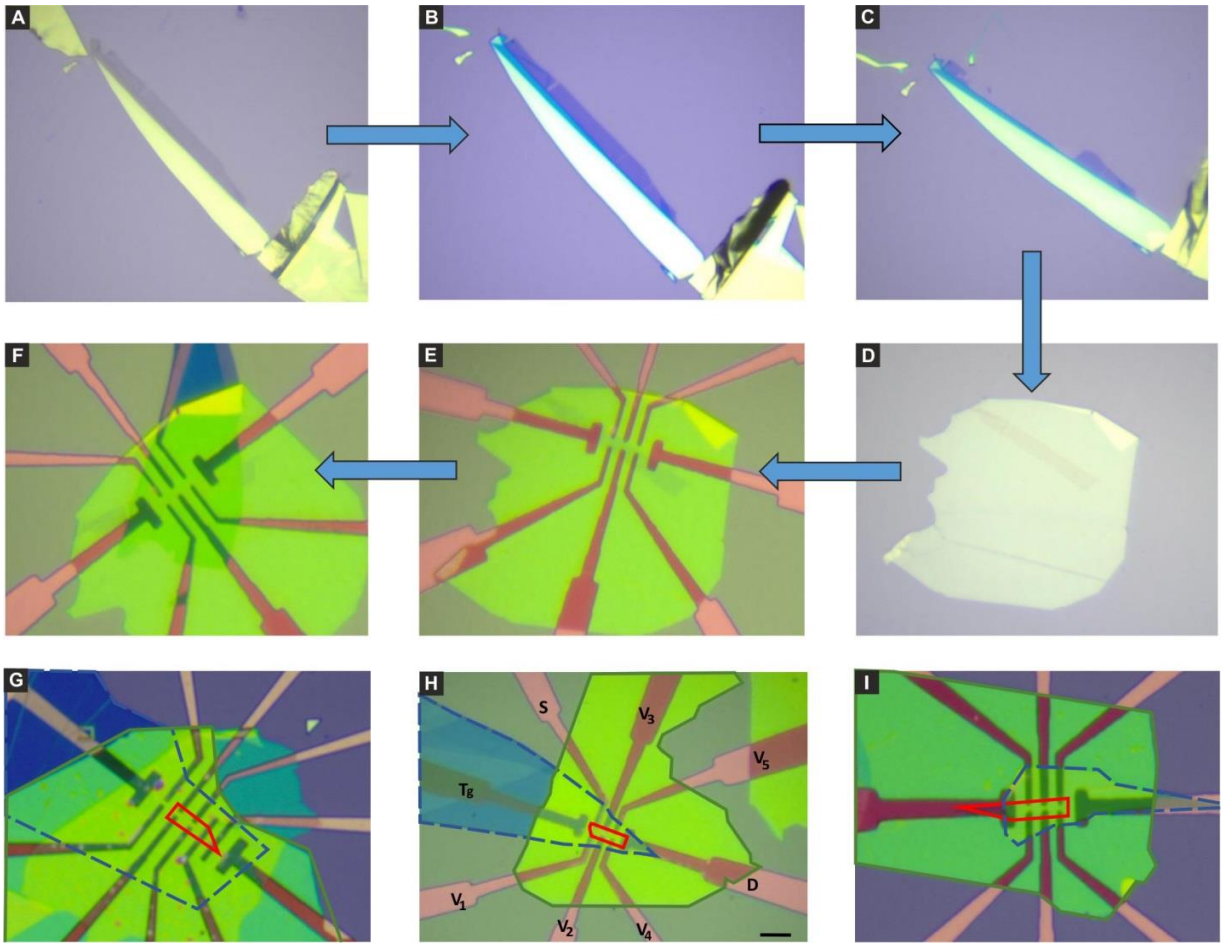

**Fig. S1 | Optical images illustrating the sequence of device fabrication steps.** Flakes of different thickness are produced by mechanical exfoliation using sticky tape. (**A**) The obtained flakes are transferred onto a PDMS stamp by pressing the tape with flakes onto the stamp. The region with weak contrast has monolayer thickness. It is about 30 μm long and is attached to 1T'-MoTe$_2$ of much larger thickness. (**B**), The flake on the PDMS stamp is transferred onto a doped Si substrate covered by a 300 nm thick thermal silicon oxide layer. A portion of the monolayer 1T'-MoTe$_2$ region is torn off from the thicker part by picking it up with an hBN film placed on top of the PDMS stamp. Panel (**C**) shows the remaining 1T'-MoTe$_2$ after tearing, while panel (**D**) is an image of the PDMS stamp with hBN as well as the monolayer 1T'-MoTe$_2$ piece. The hBN/MoTe$_2$ heterostructure was then transferred on top of another SiO$_2$/Si substrate on which a Cr(5nm)/Au(20nm) contact pattern was first deposited with the help of e-beam lithography and thermal evaporation (**E**). The flake touches the source and drain as well as the three contacts at the bottom. Finally, a graphitic layer is transferred on top of the device. It serves as the top gate (**F**). Panels (**G**)-(**I**) show the optical images of three completed devices: D2, D3 and D4. 1T'- MoTe$_2$, hBN and graphite are marked with a solid red, a solid green and a dashed blue line, respectively. The scale bar in panel (**H**) corresponds to 10 μm.

## Section S2: Sample thickness determination

In order to determine the thickness of the 1T'-MoTe$_2$ layer, we mainly use optical microscopy, Raman spectroscopy as well as the critical temperature of the superconducting transition. Optical images of layers with different thicknesses are illustrated in Fig. S2A. Layers with a thickness of up to three monolayers can be identified using the optical contrast. Raman spectra on flakes of different thicknesses are plotted in panel B. For layer thicknesses of less than four layers the position of the vibrational modes can be used to confirm the layer thickness (*55*). These Raman spectroscopy results also help to "calibrate" the optical contrast needed for a certain layer thickness. The observed modes are denoted as P1 through P9 in Fig. S2B. In previous studies, the P1, P4, P5, P6, P8 and P9 modes were identified as A$_g$ modes in the 1T' phase, whereas the P2, P3, and P7 modes were assigned as B$_g$ modes. The position of the P1 mode displays the most obvious layer thickness dependence with the peak occurring at ~86 cm$^{-1}$ for 1L, ~81 cm$^{-1}$ for 2L, ~79 cm$^{-1}$ for 3L and ~78 cm$^{-1}$ for 4L. Beyond four layers the thickness determination becomes less reliable. The data in Fig. S2B show good agreement with previously published results (*55*).

The temperature-dependence of the resistance for samples of different layer thickness is summarized in Fig. S2C. The sheet resistivity increases significantly when the layer thickness decreases from four layers to one layer. At room temperature, the resistivity is about ~600 Ω for devices with a thickness of 3 or 4 layers, ~2 kΩ for devices with two layers and ~5 kΩ for monolayer devices. For a thickness above two layers, the samples display metallic behavior and a critical temperature $T_{c,\text{onset}}$ below 3 K. The $T_{c,\text{onset}}$ increases significantly with layer thickness reduction as seen in the inset of Fig. S2C. This behavior is consistent with previously reported data in the literature (*18*, *20*). For monolayer devices, a metal-to-insulator transition is observed at lower doping density, followed by a sharp resistance drop near 7-8 K signaling the superconducting transition. Hence, the $T_{c,\text{onset}}$ can also serve as a convenient criterion to distinguish samples of different layer thickness.

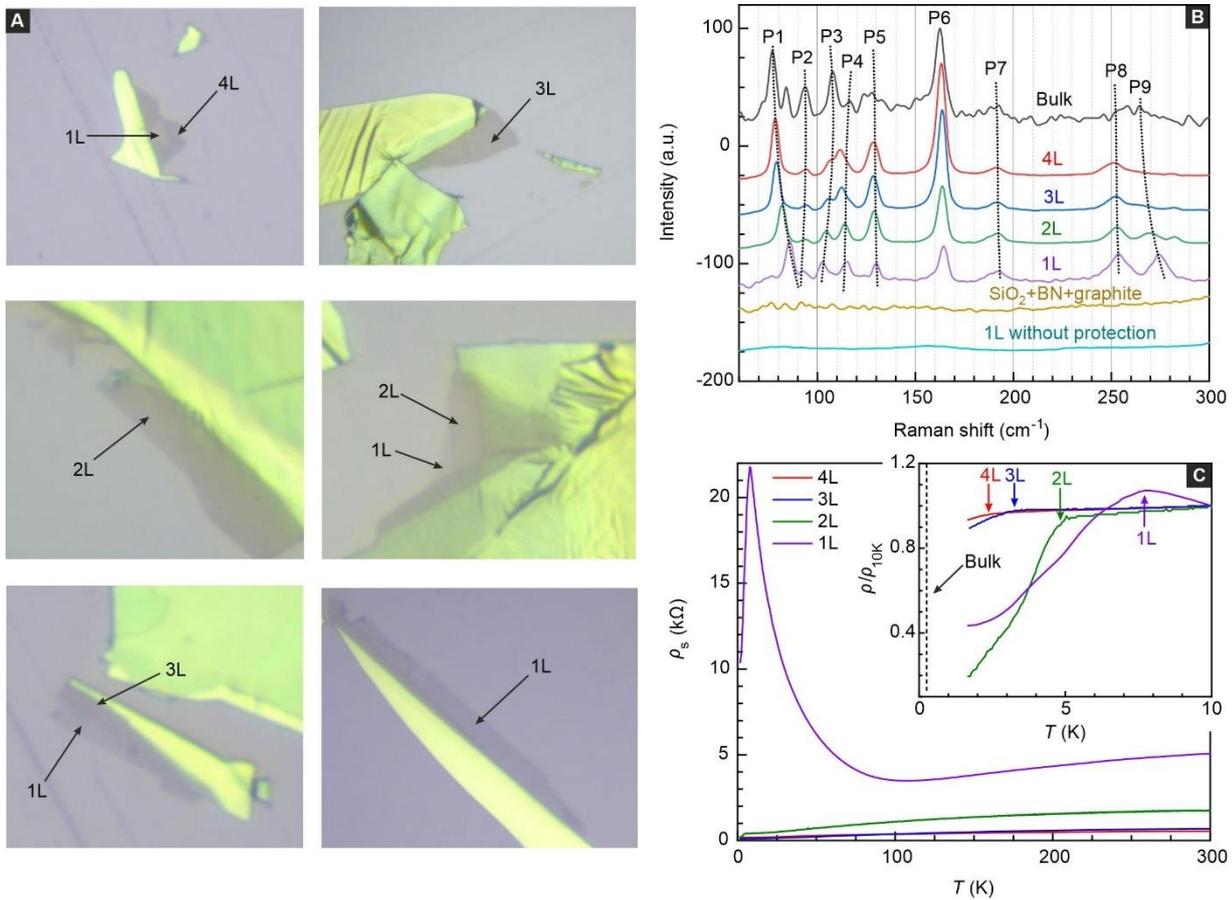

**Fig. S2 | Different methods used to obtain information about the 1T'-MoTe$_2$ layer thickness.** (**A**) Optical images of samples with a layer thickness between one and four layers. The optical contrast helps to identify the layer thickness. (**B**) Raman spectra on samples with different layer thickness from the monolayer to the bulk. The layer thickness or layer sequence of the samples used in this Raman study are marked for each trace. (**C**) Temperature-dependent resistivity of samples with 1T'-MoTe$_2$ layer thicknesses between one and four layers. The inset shows a zoom of each curve near the superconducting transition.

## Section S3: Carrier density estimates

The carrier density is estimated from a straightforward parallel plate capacitor model (*9*, *10*)

$$n_{\text{capacitance}} = \frac{c_{tg}V_{tg} + c_{bg}V_{bg}}{e} + n_0 = \frac{1}{e}\left(\frac{\varepsilon_0\varepsilon_r(\text{hBN})V_{tg}}{d_{tg}} + \frac{\varepsilon_0\varepsilon_r(\text{SiO}_2)V_{bg}}{d_{bg}}\right) + n_0.$$

Here $e$ is the electron charge, $n_0$ is the density at zero back and front gate voltage induced by disorder, $\varepsilon_0$ is the permittivity of vacuum, $\varepsilon_r(\text{SiO}_2) \sim 3.9$ and $\varepsilon_r(\text{hBN}) \sim 4$ are the relative dielectric constants of SiO$_2$ and hBN, $d_{bg} \sim 300$nm is the thickness of the SiO$_2$ thermal oxide covering the doped Si substrate, $d_{tg}$ is the thickness of the hBN encapsulating layer, $c_{bg} = \frac{\varepsilon_0 \varepsilon_r(\text{SiO}_2)}{d_{bg}}$ and $c_{tg} = \frac{\varepsilon_0 \varepsilon_r(\text{hBN})}{d_{tg}}$ are the areal capacitances for the bottom and top gate, and $V_{\text{tg}}$ and $V_{\text{bg}}$ are the top and bottom gate voltages applied to the graphite layer and the heavily doped silicon substrate, respectively. Color renderings of the longitudinal resistance $R_{xx}$ recorded across the parameter space spanned by the top and bottom gate voltages ($V_{\text{tg}}$-$V_{\text{bg}}$) at base temperature and zero magnetic field are shown in Fig. S3 for the monolayer devices D2 (panel A), D3 (panel B) and D4 (panel C).

The determination of the disorder induced density $n_0$ can in principle proceed in two different ways. In the first method it is assumed that the sample resistance reaches its maximum at zero average density. The resistance maxima should then correspond to $\frac{c_{tg}V_{tg}+c_{bg}V_{bg}}{e} + n_0 = 0$. Because the resistance maxima occur in the insulating regime, this procedure suffers from strong fluctuations in the data as is apparent in panel A-C of Fig. S3. Alternatively, it is possible to use carrier density data points extracted from Hall measurements in the high density regime and extrapolate $n_0$ from a fit of the data to the expression $n_{\text{Hall}} = \frac{c_{tg}V_{tg}+c_{bg}V_{bg}}{e} + n_0$. This method too is not without difficulties. Fig. S3D displays Hall resistance traces recorded at different pairs of the back and top gate voltages. The behavior of the Hall resistance becomes anomalous at lower density when entering the insulating regime. We would expect the Hall resistance to rise with decreasing density, yet the Hall resistance drops. This suggests that in this regime the net average density is low, but there are co-existing hole and electron populations which prevent the proper extraction of the charge carrier density from the Hall resistance formula assuming a single charge carrier type. In Fig. S3E the experimentally determined Hall densities ignoring this issue are compared with the densities determined from the capacitor model. The Hall density deviates significantly from the expected linear capacitor model behavior for densities below $1.5 \times 10^{13}$ cm$^{-2}$. This effectively constrains the useable density range for the extraction of $n_0$ and limits its accuracy. Since the longitudinal resistance increases substantially even though the drop in the Hall resistance suggests the availability of both an electron and hole population, there is apparently more to it. The insulating state may be a correlated insulator or may host edge states within the gap, as mentioned in previous publications (*9*, *20*). Due to the requirement of avoiding exposure of the MoTe$_2$ to air, it is not possible to etch the sample in a regular Hall bar geometry with well-defined contact terminals. The contact area and sample area are not well separated. All of these issues complicate the interpretation of the observed Hall resistance behavior. While both methods for extracting $n_0$ are prone to uncertainty, this has no impact on the main results. Here we list some obtained values for $n_0$ using either one of both methods. The details of the substrate are well known. For a 300 nm dry thermal SiO$_2$, $\frac{c_{bg}}{e}$ equals $7.19 \times 10^{10}$ cm$^{-2}$ for all devices. For device D2 and D3, the thicknesses of the top hBN is measured to be equal to 42 nm and 40 nm using atomic force microscopy (AFM) resulting in a $\frac{c_{tg}}{e}$ of $5.3 \times 10^{11}$ cm$^{-2}$ and $5.5 \times 10^{11}$ cm$^{-2}$, respectively. Using the first method, the intercept of the linear fit to the resistance peaks in panel A and B then yields $n_0 = 9.3 \times 10^{12}$ cm$^{-2}$ for device D2 and $n_0 = 3.4 \times 10^{12}$ cm$^{-2}$ for device D3. For device D4 gate dependent Hall measurements were used instead (Fig. S3E). For this device, we get $\frac{c_{tg}}{e} = 7.1 \times 10^{11}$ cm$^{-2}$ and $n_0 = 5.7 \times 10^{12}$ cm$^{-2}$. For device D5, $\frac{c_{tg}}{e}$ equals $8.3 \times 10^{11}$ cm$^{-2}$, while we obtain a disorder induced density $n_0$ of $1.2 \times 10^{12}$ cm$^{-2}$ using a fit to the resistance maxima (first method).

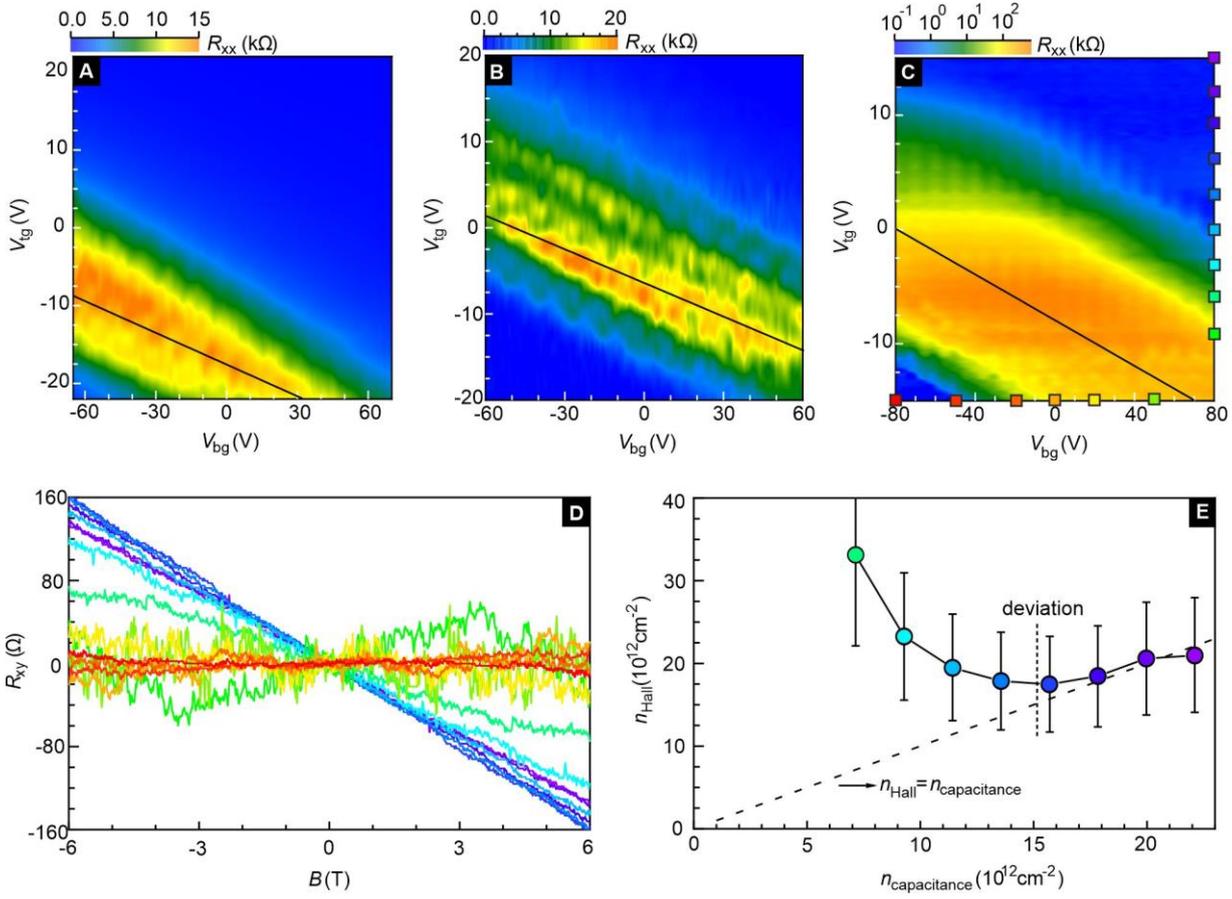

**Fig. S3 | Carrier densities extracted from the capacitance model and from Hall measurements. (A-C)** Gate dependence of the longitudinal resistance $R_{xx}$ in the $V_{tg}$-$V_{bg}$ plane for device D2, D3 and D4 from left to right. Solid black lines mark the estimated position of charge neutrality, i.e. average zero carrier density. **(D)** Hall measurements recorded on device D4 at the gate voltages marked in panel **C** using squares of the same color as each of the Hall resistance traces. **(E)** Comparison of the carrier density extracted from the Hall resistance ($n_\text{Hall}$) and the carrier density calculated from the capacitor model ($n_\text{capacitance}$). The coloring of the experimental Hall density data points again corresponds to the gate voltage pairs marked with a square of the same color in panel C.

## Section S4: Electronic phase diagram

All monolayer devices exhibit the same phase transitions from an insulating state at low density to a superconducting state for larger electron or hole doping. Fig. S4A-C illustrate the evolution of the *RT* curves below 15 K for different electron and hole densities. The resistance in the insulating regime is the highest for device D4 suggesting fewer residual charge carriers at zero average density, whereas the low resistance in device D2 in the insulating regime indicates a higher degree of inhomogeneous disorder. Panels D-F display color renditions of the resistance in the temperature versus density plane for device D2, D3 and D4. In each case, the parameter space can be subdivided in areas where the samples exhibit normal metallic behavior (NM), insulating behavior and superconducting behavior (SC). Fig. S5 shows the evolution of the *RT* curves for different densities as in Fig. S4, but for a much wider temperature range (1.6 to 200 K). The appearance of a metal-insulator transition happens in all devices at temperatures below about ~100 K as marked with a dashed black line. This heralds the opening of a bulk gap in the monolayer.

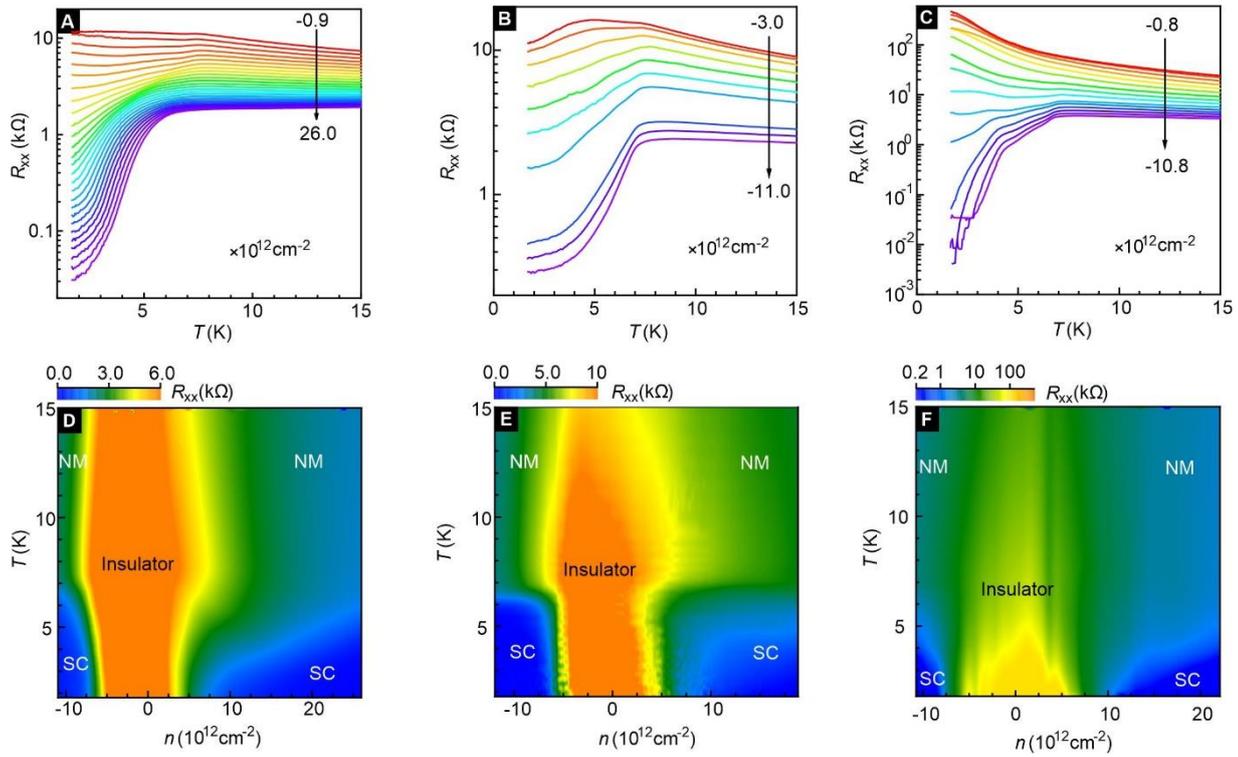

**Fig. S4 | Density dependent *RT* curves and resistance color maps as a function of density and temperature up to 15 K. (A and B)** Temperature-dependent resistance for different electron and hole carrier densities in device D2, respectively. **(C)** The same as (**A**) but for device D4 for different hole densities. **(D-F)** 2D color maps of the resistance in the plane spanned by temperature and density for device D2, D3 and D4. The different phases are marked in each diagram (insulator, normal metal - NM, superconductor - SC).

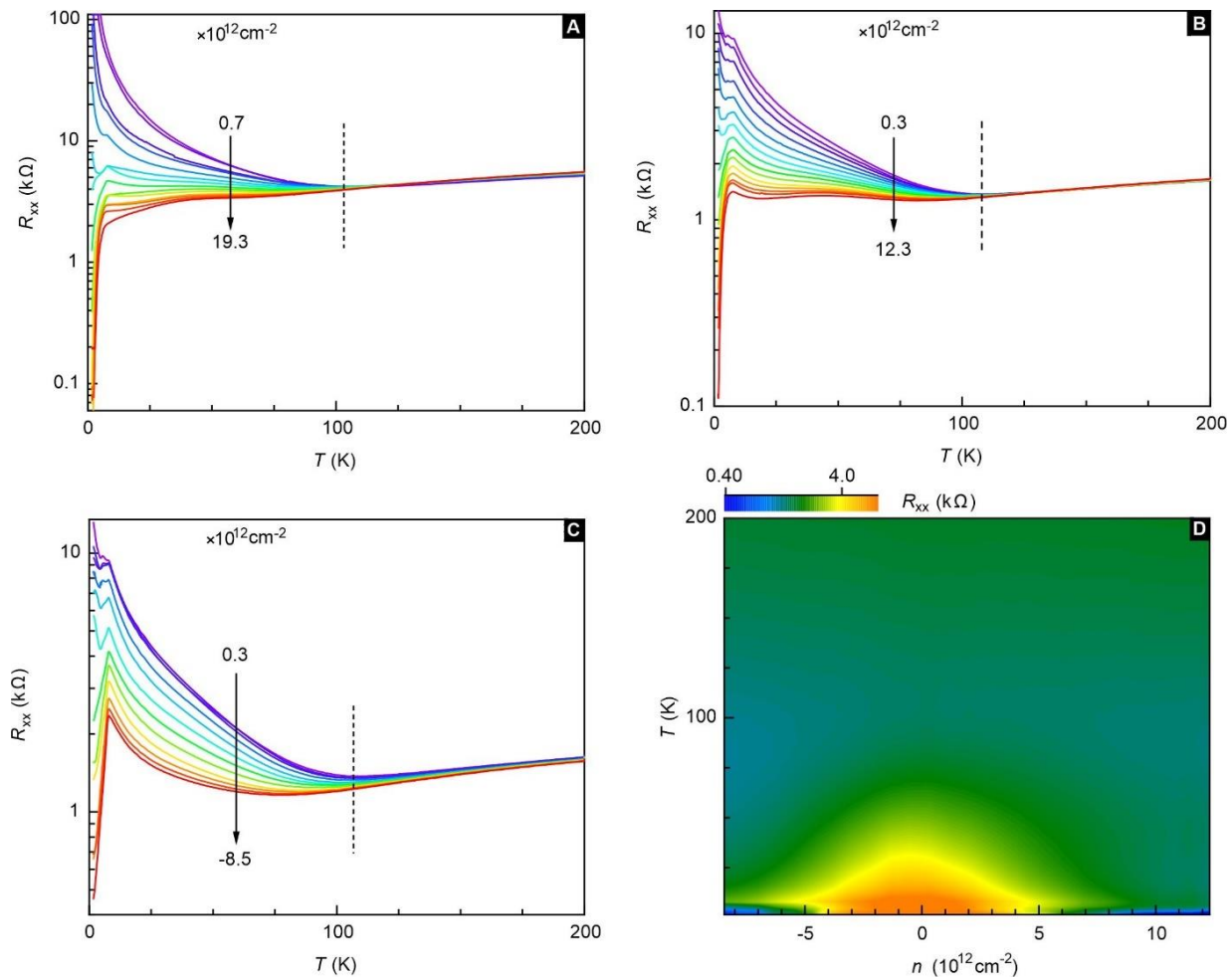

**Fig. S5 | Density dependent *RT* curves and resistance plot as a function of density and temperature up to 200 K. (A)** Temperature-dependent resistance traces for different electron densities recorded on device D4. The dashed black line marks the temperature at which a metal-to-insulator transition appears when tuning the density. **(B and C)** Same as **(A)** but for device D5 for different electron **(B)** and hole **(C)** densities. **(D)** 2D color map of the resistance in the parameter plane spanned by the density and temperature for device D5.

## Section S5: Density and bias current dependence of the differential resistance

Fig. S6 displays the behavior of the differential resistance $dV/dI$ as a function of the bias current and the carrier density for three different monolayer samples: D2, D3 and D4. The top panels are waterfall plots whereas the bottom panels are color renderings of the differential resistance in the dc bias current and carrier density plane.

The large $dV/dI$ peaks for zero bias current reflect the insulating state (Ins) at low densities. The peak value strongly depends on the device, which we attribute to the significant variation in the residual spatially inhomogeneous carrier density. As the electron (e) or hole (h) density is raised, the peak in the differential resistance at zero bias current vanishes and the differential resistance drops to zero at low bias current instead. This region of zero differential resistance expands with increasing electron or hole density and the critical current where the differential resistance becomes non-zero moves to higher bias current. The density dependence of the critical current is not symmetric when comparing electron and hole doping. This presumably reflects the lack of symmetry in the band structure.

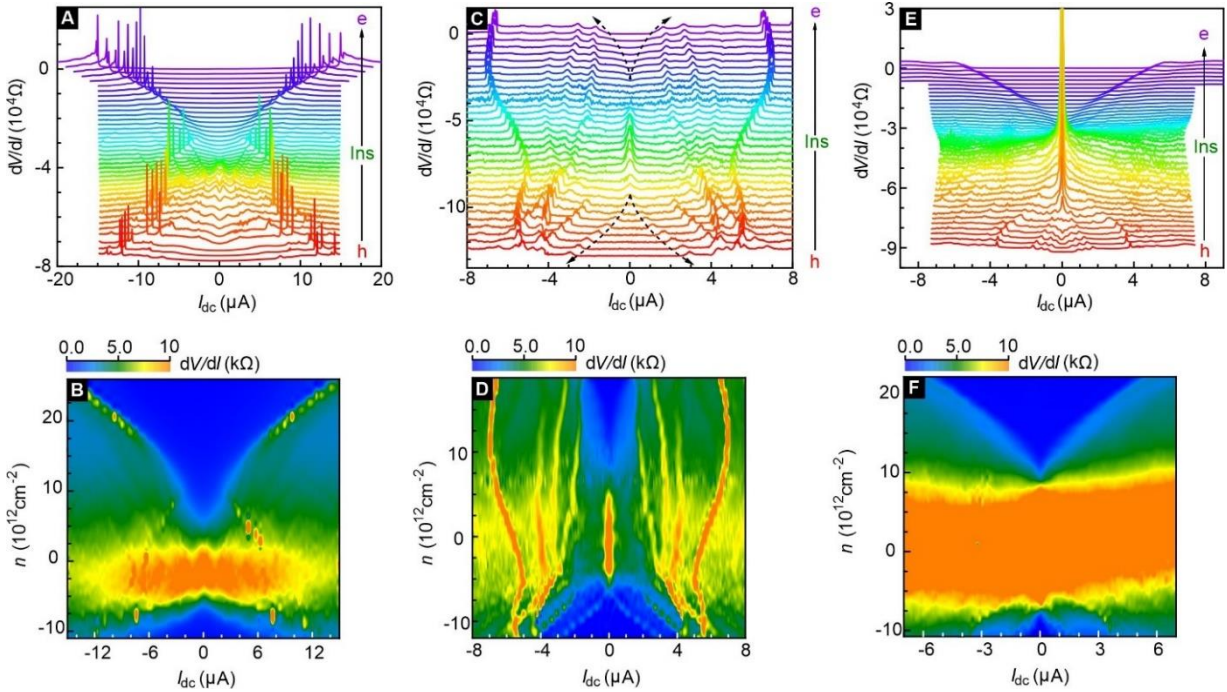

**Fig. S6 | Differential resistance $dV/dI$ as a function of dc bias current and charge carrier density. (A)** Waterfall plot of the differential resistance recorded on device D2 at $T = 1.7$ K and $B = 0$ T for densities ranging from $-1.10\times10^{13}$ cm$^{-2}$ to $2.60\times10^{13}$ cm$^{-2}$. Each curve is offset by 2 k$\Omega$. **(B)** 2D color plot of the differential resistance data shown in panel **(A)**. **(C and D)** The same as **(A)** and **(B)** but for device D3 with densities ranging from $-1.20\times10^{13}$ cm$^{-2}$ to $1.88\times10^{13}$ cm$^{-2}$. The offset in **(C)** is 4 k$\Omega$. **(E and F)** The same as **(A)** and **(B)** but for device D4 with densities ranging from $-1.08\times10^{13}$ cm$^{-2}$ to $2.21\times10^{13}$ cm$^{-2}$. Each curve is offset by 2 k$\Omega$ in **(E)**.

## Section S6: Estimate of the bulk gap

It is possible to estimate the size of the bulk gap Δ by performing an Arrhenius plot analysis of the temperature dependent resistance data in the regime where the sample exhibits insulating behavior: $R_{xx}(T) \sim \exp\left(\frac{\Delta}{2k_BT}\right)$ (56). The analysis should be performed at gate bias points where the chemical potential is located inside the bulk gap and the average density is low. Inhomogeneous disorder, sufficiently large to generate a landscape of electron and hole puddles, will likely cause an underestimation of the gap size. Hence, higher quality samples with less disorder are more suitable for such an analysis. In Fig. S7A an Arrhenius fit is performed on the resistance data obtained on device D4 with the highest quality at a density of $7\times10^{11}$ cm$^{-2}$. The main graph plots the temperature dependent resistance on a logarithmic scale, whereas the inset displays the Arrhenius fit (solid red line) to the conductance:

$$\frac{1}{R_{xx}} = G_{xx}(T) \sim \exp\left(-\frac{\Delta}{2k_BT}\right),$$

focusing on the high-temperature regime between 20 and 100 K. This yields a value for the bulk gap of ~7.3meV. The Arrhenius fit deviates significantly at temperatures below 20 K. This temperature regime is described well by the expression for variable-range hopping (VRH) in a 2D system with localization $[G_{xx}(T) \sim \exp\left(-\left(\frac{T_0}{T}\right)^{\frac{1}{3}}\right)]$, as seen in the inset of Fig. S7A (solid black line). Even if so, other potential mechanisms can't be totally ruled out (57). In another device D5, the estimated bulk gap is ~7.0 meV for a net density of $3\times10^{11}$ cm$^{-2}$ (Fig. S7B). This is close to the gap value obtained on device D4 in panel A.

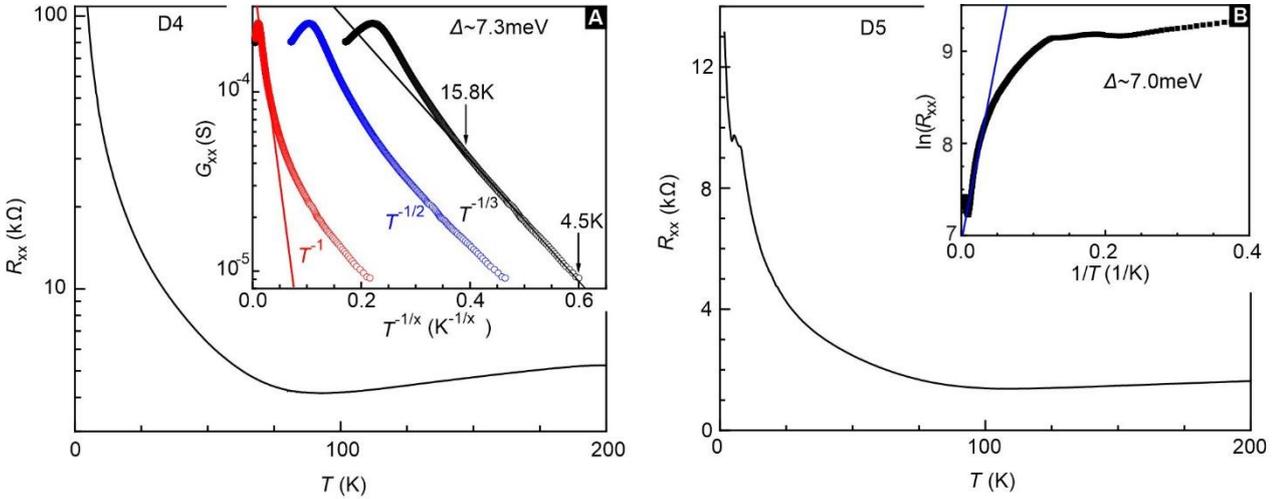

**Fig. S7 | Estimate of the bulk gap for samples with different degrees of inhomogeneous disorder. (A)** Temperature-dependent resistance for device D4 in the insulating regime at a density of $7\times10^{11}$ cm$^{-2}$. The inset shows an Arrhenius fit for temperatures from 20 to 100 K (red line) and a fit assuming variable range hopping in the temperature interval from 4 to 20 K (black line with $T^{-1/3}$). The blue curve is shown for the sake of comparison with a $T^{-1/2}$ dependence. **(B)** The same as **(A)** but for device D5 at a net density of $3\times10^{11}$ cm$^{-2}$.

## Section S7: Density dependence of the resistance at different temperatures and fields

Fig. S8 plots the density dependence of the resistance for different temperatures and magnetic fields for devices D2, D4 and D5. By sweeping the gate voltage, the chemical potential is tuned continuously from the valence band into the conduction band. This results in ambipolar transport behavior with insulating behavior in between when the average density is small. The highest resistance in the insulating regime is strongly sample-dependent. It ranges from ~$10^4$ ohms to ~$10^5$ ohms, as seen in panels A, C and E of Fig. S8. There is no clear evidence for the existence of topological edge states. Such edge states would cause a resistance peak with a quantized value of about $h/2e^2$. Resistances are much larger. Moreover, they are sample-dependent (*31*, *32*). Panels B, D and F plot the magnetic field dependence of the resistance for samples D2, D4 and D5. The resistance in the insulating regime increases gradually with magnetic field. This contradicts the expected behavior for quantum spin Hall edge states when time-reversal symmetry gets broken in the presence of a small magnetic field. This lack of experimental signatures for edge states may be the result of remaining disorder and the inevitably associated scattering. Short-channel devices are more appropriate to mitigate at least to some extent the influence of disorder for edge state related transport (*58*, *59*).

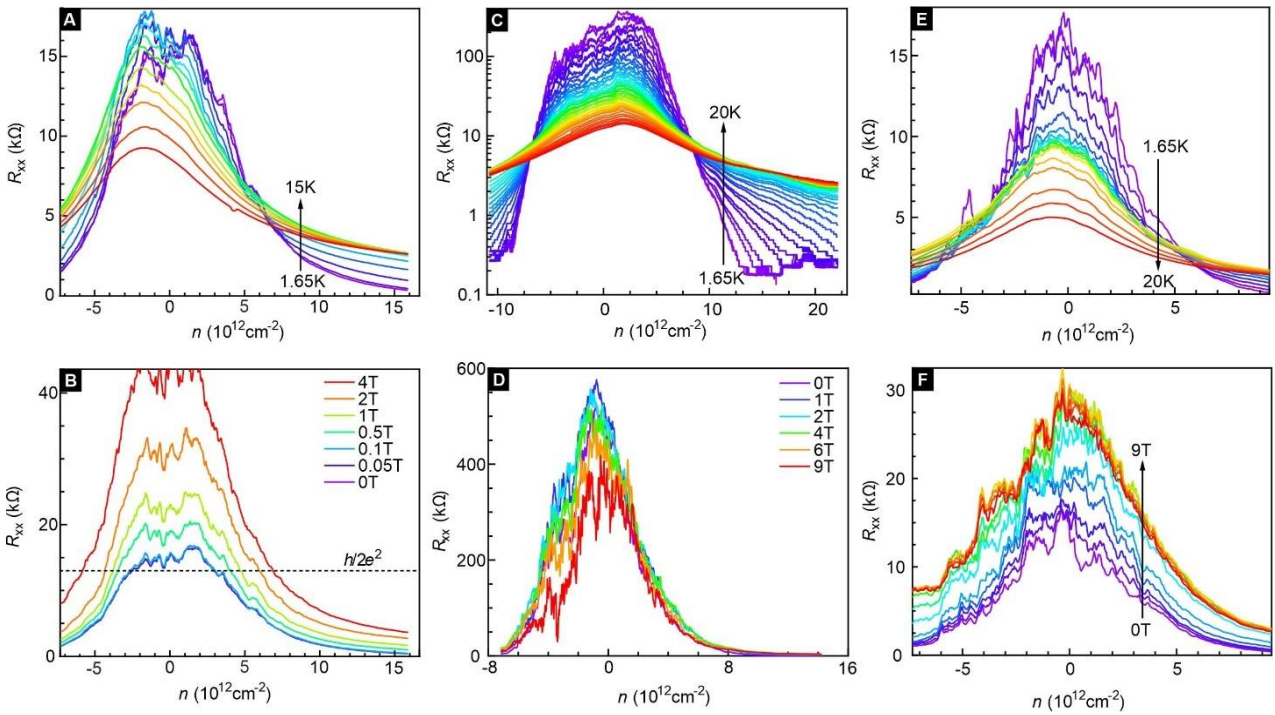

**Fig. S8 | Density dependence of the resistance at different temperatures and magnetic fields.** (**A**) Density dependent longitudinal resistance data for device D2 in the absence of a magnetic field and with temperature as a parameter. The temperature range is 1.65 to 15 K in steps of about 1 K. (**B**) Same as in (**A**) but at a fixed temperature of 1.65 K and with the magnetic field as the parameter. (**C** and **D**) Same as (**A**) and (**B**) but for device D4. The covered temperature range is from 1.65 to 20 K in approximately 0.25 K steps. (**E** and **F**) Same as (**A**) and (**B**) but for device D5. The temperature range from 1.65 to 20 K is covered in 1 K steps (**E**). The magnetic field varies from 0 to 9 T in 1T steps.

## Section S8: The low density regime

In the low density regime, the resistance behaves as an insulator for temperatures typically down to about 7-8 K marked by the gray area in the exemplary density-dependent *RT* traces recorded on device D1, D2 and D4 in Fig. S9A-C. The kink in the traces in this temperature regime signals the onset of the superconducting transition and does not change significantly when varying the doping level. This is reminiscent of a superconductor-to-insulator transition in granular films of various metals (*28*, *60*). The behavior is attributed to the existence of isolated superconducting puddles in an insulating background (*61*), as shown in a cartoon like fashion in the inset of Fig. S9C. This landscape originates from the spatial density inhomogeneity across the sample where areas of high-density remain conducting. Upon reducing the temperature below the transition temperature $T_{c,\text{onset}}$ Cooper pairs form within these conducting puddles and cause at least a kink or a pronounced downturn of the resistance. The carrier density averaged across the entire sample area is still very small, but inside these puddles the carrier density should be on the order of ~$10^{12}$ cm$^{-2}$ and can be tuned by changing the applied gate voltages. The transition temperature $T_{c,onset}$ does not or depends only weakly on the applied gate voltage even down to lowest average sample densities as can be seen in Fig. S9D. The traces plotted in panel A and B for higher average carrier densities indicate that the resistance continues to decrease as temperature is lowered, but the resistance values remain non-zero and large. This suggests that there is only weak link coherence among the puddles and overall superconductivity cannot be achieved.

The existence of superconducting puddles is also corroborated in temperature dependent resistance data in the presence of a perpendicular magnetic field. When the magnetic field is increased, the superconducting puddles should gradually convert into the normal state at relatively high fields (> 2T) and the insulating behavior should then persist also for temperatures below 7 K. This is indeed observed in perpendicular field data plotted in Fig. S10. The low-temperature resistance reaches values as high as ~$10^5$ ohms when the superconducting puddles are entirely quenched in high magnetic fields.

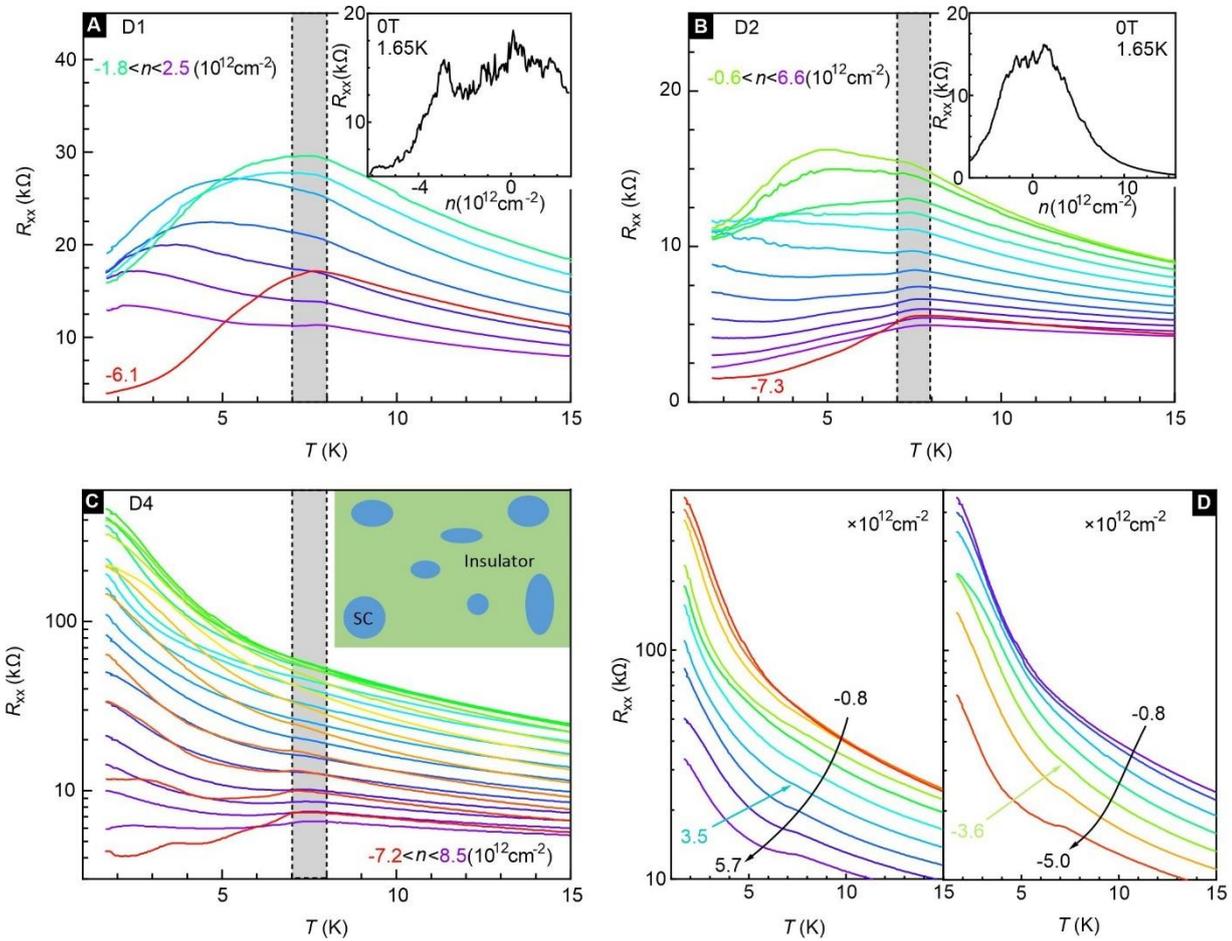

**Fig. S9 | Superconducting puddles in the insulating regime.** (**A**) $RT$ curves recorded on device D1 for varying carrier densities (red curve: -6.1×$10^{12}$ cm$^{-2}$, green to purple curves: -1.8<$n$<2.5×$10^{12}$ cm$^{-2}$). The gray area marks the common transition-temperature region. The inset shows the gate-dependent resistance at 1.65 K and zero magnetic field. (**B**) Same as (**A**) but for device D2 (red curve: -7.3×$10^{12}$ cm$^{-2}$, green to purple curves: -0.6<$n$<5.3×$10^{12}$ cm$^{-2}$. (**C**) Same as (**A**) but for device D4 (-7.2<$n$<8.5×$10^{12}$ cm$^{-2}$). The inset displays the schematic of the superconducting puddles. (**D**) Same as (**C**) but divided into two panels for electron (left panel) and hole (right panel) doping, respectively.

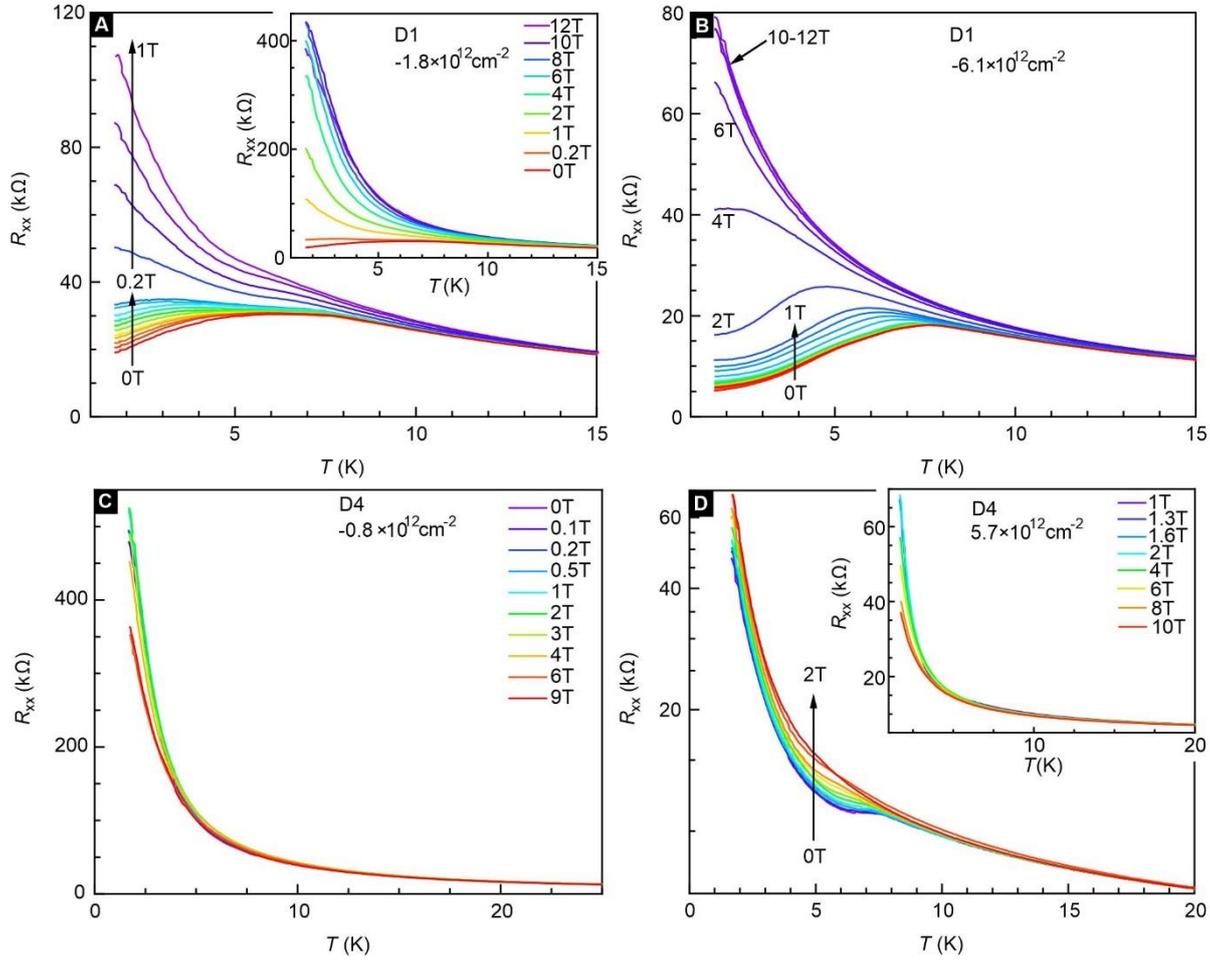

**Fig. S10. Magnetic field-dependent behavior in the insulating regime.** (**A**) *RT* curves recorded in the presence of different perpendicular magnetic fields (from 0 to 1 T) on device D1 for an average hole density of $-1.8 \times 10^{12}$ cm$^{-2}$. The inset displays data recorded at higher magnetic fields (0-12 T). (**B**) Same as (**A**) but for a hole density of $-6.1 \times 10^{12}$ cm$^{-2}$. (**C**) Same as (**A**) but for device D4 for a hole density of $-8 \times 10^{11}$ cm$^{-2}$. (**D**) The same as (**C**) but for an electron density of $5.7 \times 10^{12}$ cm$^{-2}$.

## Section S9: Density dependence of the superconducting parameters

In Fig. S11, we summarize the density dependence of some key sample and superconductivity parameters, such as the superconducting transition temperature $T_{c,0}$, the out-of-plane critical field $B_{c2,\perp}$, the BCS coherence length $\xi_0$ and the mean free path $l_{mfp}$. Within the density range we can cover, the superconductivity strengthens with density, since both $T_{c,0}$ in Fig. S11A and $B_{c2,\perp}$ in Fig. S11B continue to increase with increasing density without observing a maximum that would signal optimal doping. In order to judge whether the superconductor is in the dirty limit ($\xi_0 \gg l_{mfp}$) or the clean limit ($l_{mfp} \gg \xi_0$), the BCS coherence length is compared with the mean free path for different densities (*33, 34*). The mean free path follows from the Drude model: $l_{mfp} = h/(e^2 \rho_s \sqrt{g_s g_v \pi n_e})$ with $\rho_s$ the sheet resistivity, $n_e$ the net charge carrier density. For electron densities, $g_s = g_v = 2$ to account for both the spin and valley degeneracy. For hole doping, $g_s = 2$ and $g_v = 1$ (*10, 20*). We can obtain the BCS coherence length from the in-plane Ginzburg-Landau (GL) coherence length using the relation $\xi_0 = 1.35\, \xi_{GL}(0K)$. The in-plane coherence length $\xi_{GL}(0K)$ is calculated using the expression $B_{c2,\perp}(0K) = \frac{\phi_0}{2\pi \xi_{GL}^2}$, where $\phi_0$ stands for the superconducting flux quantum $h/2e$ and $B_{c2,\perp}(0K)$ is the out-plane critical field at 0 K. The latter is obtained from a linear extrapolation of temperature dependent measurements of the out-of-plane critical field. A comparison of $l_{mfp}$ and $\xi_0$ is displayed in Fig. S11C for device D4 and in Fig. S11D for device D2. Since $\xi_0$ is always larger than $l_{mfp}$, both devices are in the dirty limit. When decreasing the carrier density down to the insulating regime, $\xi_0$ increases, whereas $l_{mfp}$ decreases. Hence, the devices are located even deeper into the dirty regime with decreasing density.

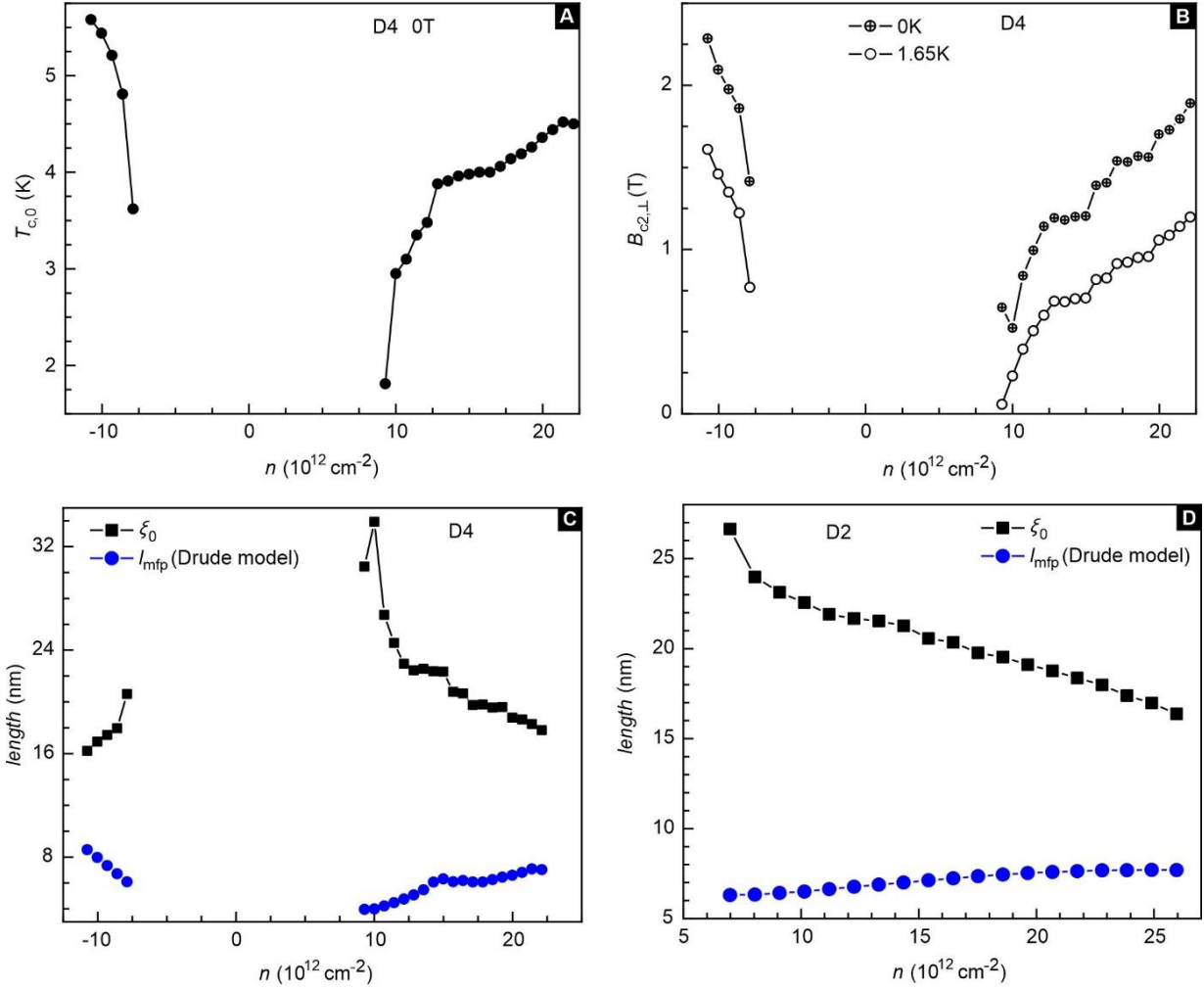

**Fig. S11 | Density dependence of the superconducting parameters.** Density dependence of $T_{c,0}$ (**A**) and $B_{c2,\perp}$ at 1.65 K and 0K (**B**) for device D4. $B_{c2,\perp}(0K)$ is obtained from a linear extrapolation of temperature dependent $B_{c2,\perp}$ measurements. (**C**) Comparison of the estimated in-plane BCS coherence length $\xi_0$ (black symbols) and mean free path $l_{mfp}$ (blue symbols) for device D4. (**D**) The same as (**C**) but for device D2.

## Section S10: Fit of the KLB model to the $B_{c2,\parallel} - T_c$ data

Since the superconductivity in our samples occurs in the dirty regime, spin-orbit scattering can be a potential mechanism enhancing the in-plane critical field $B_{c2,\parallel}$ with respect to the BCS Pauli limit. This is captured by the KLB theory. It yields the following expression for the in-plane field dependence of the critical temperature (*10, 40*)

$$\ln\left(\frac{T_c}{T_{c,0}}\right) = \psi\left(\frac{1}{2}\right) - \psi\left(\frac{1}{2} + \frac{g^2 \mu_B^2 B_{c2,\parallel}^2}{2\hbar \tau_{SO}^{-1}} \cdot \frac{1}{2\pi k_B T_c}\right).$$

Here, $\tau_{SO}$ is the spin-orbit scattering time, $g$ is the electronic $g$-factor and $\mu_B$ is the Bohr magneton. By fitting this expression to the experimental data of device D2, we obtain a $\tau_{SO}$ of ~170 fs for an electron density of $2.38 \times 10^{13}$ cm$^{-2}$ (Fig. S12A) and ~120 fs for a hole density of $-1.10 \times 10^{13}$ cm$^{-2}$ (Fig. S12B). To verify the applicability of the KLB model, it is important to confirm that the transport scattering time $\tau_{tr}$ is smaller than $\tau_{so}$. The transport scattering time can be estimated from the Drude model using the expression $\tau_{tr} = l_{mfp}/v_F$. Here, the Fermi velocity $v_F = \hbar k_F/m^* = \frac{\hbar}{m^*}\sqrt{\frac{4\pi n_e}{g_s g_v}}$. The effective mass $m^*$ is taken as $0.37 m_e$ (20). We obtain a transport scattering time $\tau_{tr}$ of ~20 fs for the electron case illustrated in Fig. S12A and ~60 fs for the hole case in Fig. S12B. Since $\tau_{tr} < \tau_{SO}$, the KLB model is indeed applicable and spin-orbit scattering may also contribute to an enhancement of the in-plane critical field, in addition to the increase associated with a strengthening of the superconducting gap due to a pair interaction enhancement as discussed in the main text and section S12 (see below).

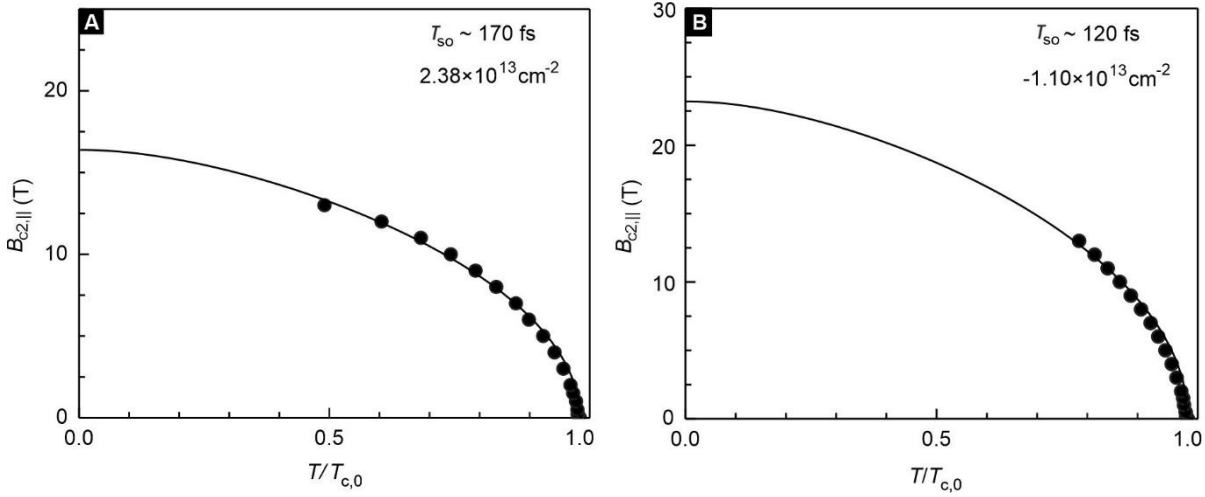

**Fig. S12 | Fit of the KLB model to the $B_{c2,\parallel} - T_c$ data of device D2.** (**A**) $B_{c2,\parallel}$ as a function of temperature (circular data points) for an electron density of $2.38 \times 10^{13}$ cm$^{-2}$. The solid black line is the fit of the KLB model. (**B**) The same as (**A**) but for a hole density of $-1.10 \times 10^{13}$ cm$^{-2}$.

### Section S11: Magnetic field induced superconductor-to-metal transition

In Fig. S13A, a superconductor-to-metal transition is observed in device D2 at an electron density of $2.60 \times 10^{13}$ cm$^{-2}$ when applying a perpendicular magnetic field larger than ~3 T. This transition is also confirmed in the magnetic field dependence of the differential resistance shown in Fig. S13B. A dip at low dc bias current changes into a peak near ~3 T. Fig. S13C illustrates the magnetoresistance for different temperatures from 1.65 K to 9 K. The isotherms show crossover points ($B_c$) within a small magnetic field region near ~3 T. The inset in Fig. S13C shows a close-up view of this crossover area. The width of the crossover region changes with density as illustrated in Fig. S14. It shrinks with increasing density. This observation suggests a magnetic field induced quantum phase transition in this 2D superconductor.

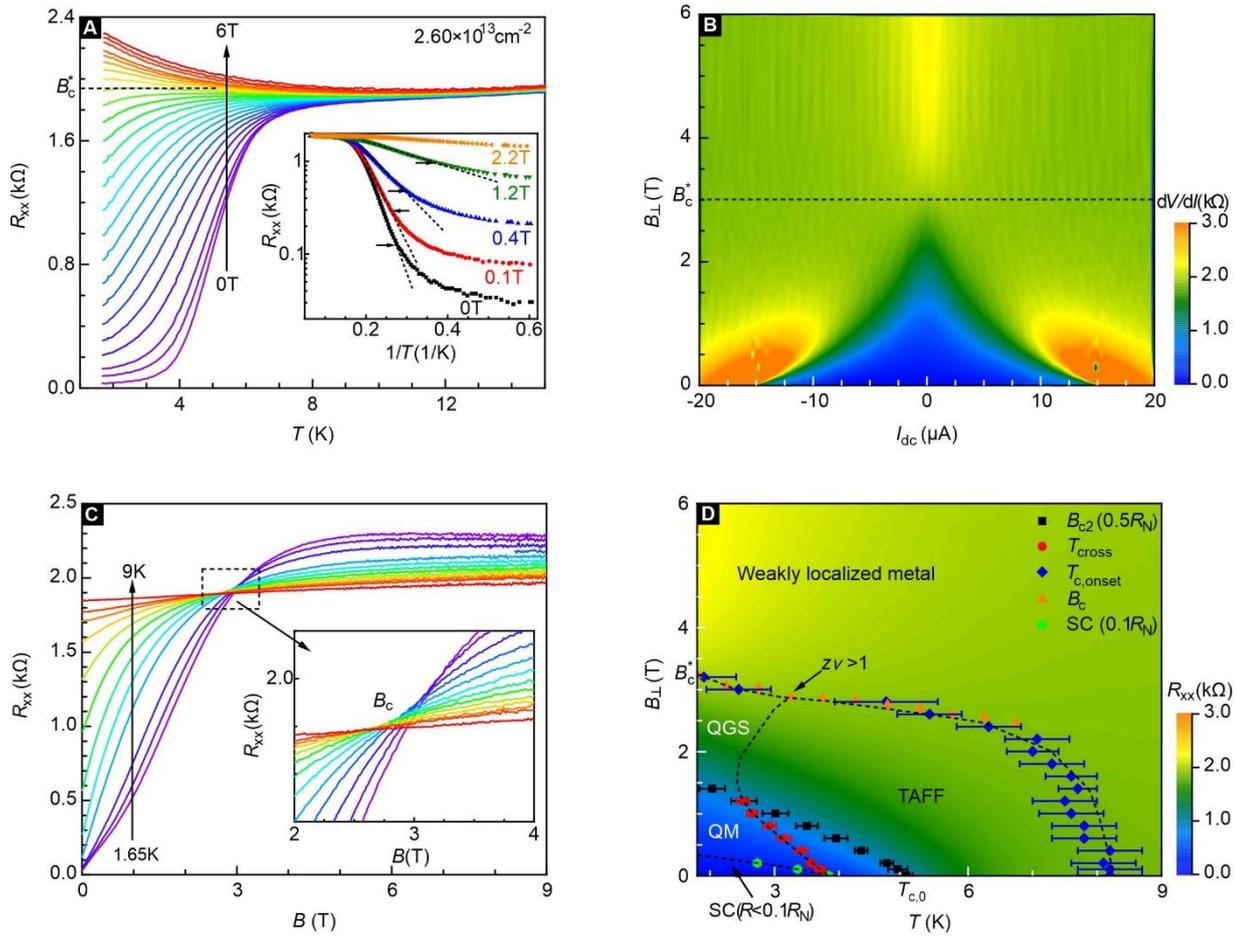

**Fig. S13. Magnetic field driven quantum phase transition in the superconducting regime**. All data were taken on device D2 in the superconducting state for an electron density of $2.60 \times 10^{13}$ cm$^{-2}$. (**A**) Temperature-dependent resistance at different perpendicular magnetic fields. The inset shows Arrhenius plots of the resistance to highlight the thermally activated behavior. The arrows mark the transition from thermally activated flux flow to a quantum metal state. (**B**) d$V$/d$I$ versus dc bias current at different magnetic fields at 1.65 K. (**C**) The magnetoresistance curves at various temperatures from 1.65 to 9 K. The inset shows the crossover region. (**D**) $B$-$T$ phase diagram for the following regimes: weakly localized metal, quantum Griffiths state (QGS), thermally activated flux flow (TAFF) and quantum metal. The dashed lines and symbols mark phase boundaries. The different symbols correspond to different criteria to identify the various phases. They are described in more detail in the text.

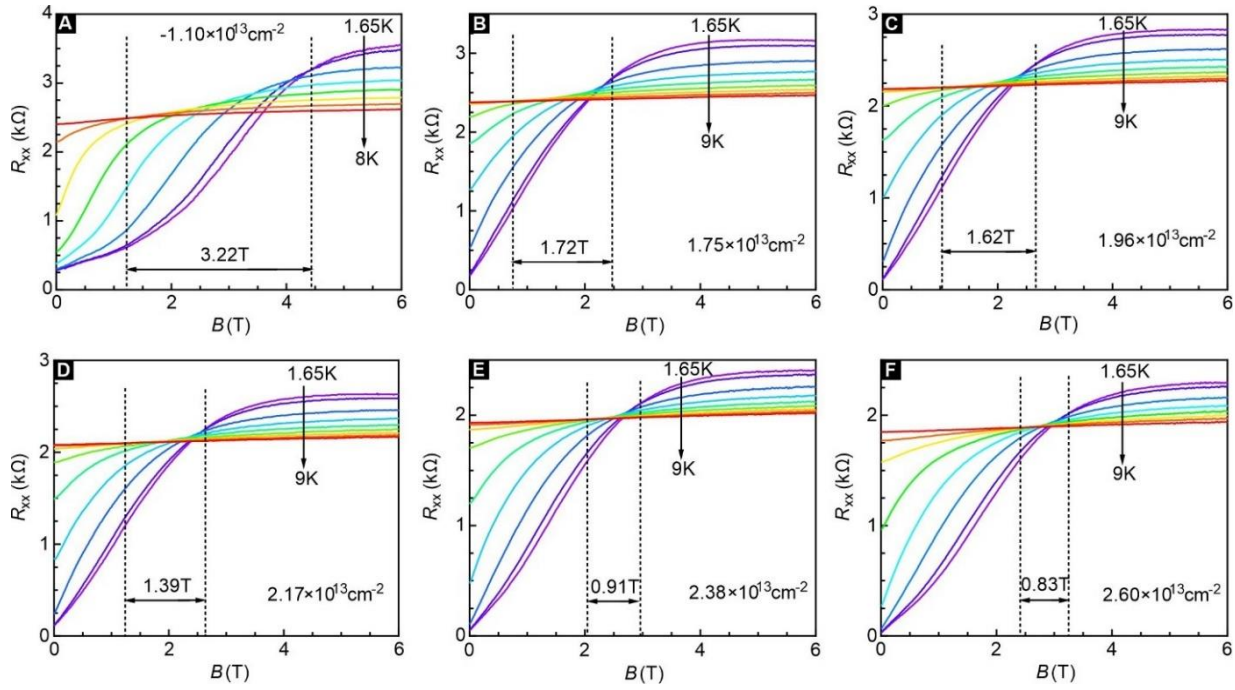

**Fig. S14. Gate dependence of the magnetoresistance in the crossover region at different charge carrier densities. (A-F)** Temperature-dependent magnetoresistance recorded on device D2 for temperatures ranging from 1.65 K to 9 K and for a hole density of -1.10 × 10$^{13}$ cm$^{-2}$ (**A**) as well as electron densities of 1.75 × 10$^{13}$ cm$^{-2}$ (**B**), 1.96 × 10$^{13}$ cm$^{-2}$ (**C**), 2.17 × 10$^{13}$ cm$^{-2}$ (**D**), 2.38 × 10$^{13}$ cm$^{-2}$ (**E**) and 2.60 × 10$^{13}$ cm$^{-2}$ (**F**). The dashed lines in each panel mark the approximate width of the crossover region.

In 2D superconducting systems, fluctuations, disorder, and finite size effects play a particularly important role. They can have a strong impact on the BKT transition, the superconductor-insulator and superconductor-metal transition. For conventional phase transitions, the dynamical critical exponent $z$ only exhibits a singularity at the quantum critical point of the phase diagram itself, while it remains constant near the quantum critical point. However, in some cases a collection of rare ordered regions can completely destroy the sharpness of the phase transition and this may be the origin of the appearance of a crossover region rather than a single crossover point in Fig. S14. This is reminiscent of a Griffiths singularity. Such a singularity is frequently observed for weakly disordered quantum phase transitions (*30*, *62*). It has been well studied for 3D systems and more recently multiple experimental observations in 2D crystalline superconductors with weak pinning potentials were reported (*2*, *29*, *63*). In these studies, multiple critical points were observed and through a finite size scaling analysis, a divergent behavior of the dynamical critical exponent, as expected for a Griffiths singularity, was confirmed. The exponent follows from the power law $z\nu \sim C(B_c^* - B)^{-0.6}$, where $C$ is a constant, $B_c^*$ is the infinite randomness critical point and $\nu$ is the static critical exponent. The theoretically predicted value for $\nu$ for a superconductor to metal transition is 0.5 (*29*). The divergence implies activated scaling with a continuously varying dynamical critical exponent when approaching the infinite-randomness quantum critical point. In our samples, when increasing the charge carrier density, the 2D superconductor becomes more homogeneous and more robust against fluctuations and disorder. This is likely responsible for the shrinking of the crossover region. Fig. S15 summarizes a typical scaling analysis. In order to extract the dynamical critical exponent $z$, a set of $R$ versus $B$ curves were recorded at different temperatures from 1.7 K to 7.5 K on device D2 for an electron density of 2.60 × 10$^{13}$ cm$^{-2}$. There are multiple crossover points ($B_c$). The finite size scaling law for the isotherms for a proximate crossover point can be expressed as $R(B,T) = R_c \cdot F(|B - B_c|(T/T_0)^{-1/z\nu})$, where $F$ is an arbitrary function obeying the condition that $F(0) = 1$. $R_c$ and $B_c$ are the critical resistance and critical magnetic field, respectively. $T_0$ is the lowest temperature for the $R$ vs. $B$ curves used in the scaling analysis. By adjusting the value of $z\nu$, curves for different temperatures can be made to collapse onto a single curve (*9*, *29*). An example of this analysis for a narrow range of temperatures between 1.7 K and 1.9 K is shown in Fig. S15B. Fig. S15C plots a quantity that quantifies how well the curves collapse onto each other for a given $z\nu$. The minimum corresponds to the $z\nu$ value that fits best. Fig. S15D displays the optimized value of $z\nu$ as a function of the magnetic field and indeed $z\nu$ rises rapidly with increasing field consistent with divergent behavior. The corresponding change with temperature is included as an insert in Fig. S15C. The data points of Fig. S15D match best the divergent power-law relation $z\nu \sim C(B_c^* - B)^{-0.6}$ and this yields a $B_c^* \sim 3.23$ T. These results provide evidence for a Griffiths singularity at $B = B_c^*$.

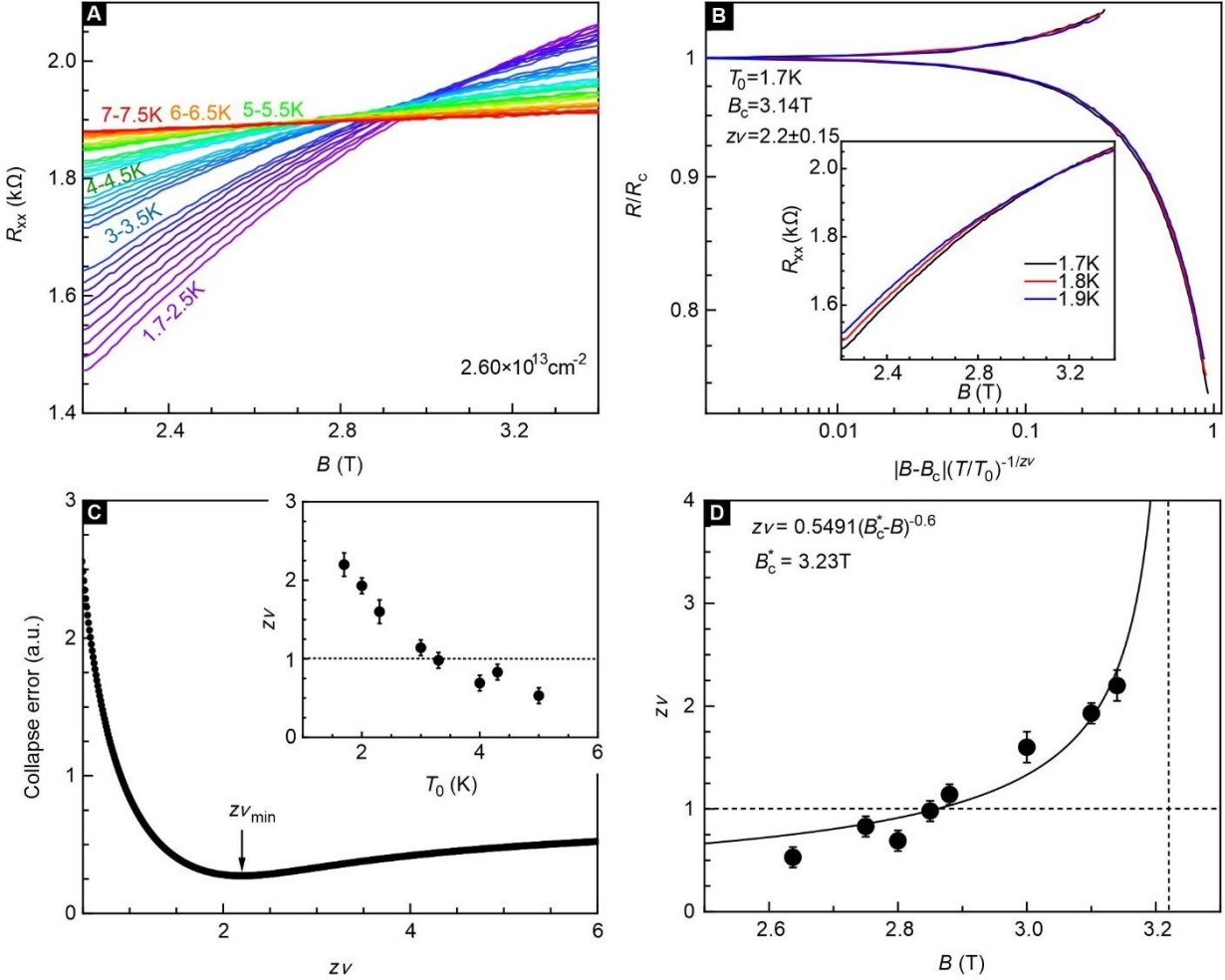

**Fig. S15. Scaling analysis in the context of a quantum Griffiths singularity.** (**A**) A close-up view of multiple crossover points for the $R$ versus $B$ curves in the temperature interval from 1.7 K to 7.5 K. These data sets were recorded on device D2 for an electron density of $2.60 \times 10^{13}$ cm$^{-2}$. (**B**) Scaling analysis of the isotherms for the temperature interval between 1.7 K and 2.0 K using the relationship $R/R_c \sim |B - B_c|(T/T_0)^{-1/z\nu}$. Here, $T_0$ refers to the lowest temperature. It equals 1.7 K. The crossover point is taken as $B_c = 3.14$ T. The inset shows some exemplary $R$ versus $B$ curves in this temperature range. (**C**) The "collapse error" as a function of $z\nu$ to identify the best fit value for $z\nu$. The inset shows the fitted $z\nu$ as a function of temperature. (**D**) Divergent behavior of $z\nu$ as a function of magnetic field $B$. The solid curve is a fit based on the activation scaling law, $z\nu \sim 0.5491(B_c^* - B)^{-0.6}$. The horizontal dashed line marks $z\nu = 1$ and the vertical line corresponds to $B_c^* = 3.23$ T.

Finally, in Fig. S13D we show a preliminary phase diagram for the transition of the superconducting ground state to the metallic state. Five different criteria are used to determine phase boundaries. The black squares correspond to the out-of-plane upper critical field, $B_{c2,\perp}(T)$. It is defined here as the field where the resistance has dropped to 50% of the normal state resistance using the data in Fig. 13A. The red circles mark the transition from the regime of thermally activated flux flow (TAFF) to the quantum metal (QM) regime. In experiment this transition is signaled by a deviation from thermally activated behavior as described by the Arrhenius expression. This transition has been marked with arrows in the inset of Fig. 13A. The blue diamonds mark the onset of superconductivity ($T_{c,\text{onset}}$). The orange triangles mark the crossover point of the $R$ versus $B$ curves recorded at neighboring temperatures (see the inset of Fig. S13C). Green hexagons signal the transition to the true superconducting state when $R < 0.1 R_N$ and $T < T_{\text{BKT}}$ at low magnetic field. With increasing magnetic field, the true superconducting state converts to a weakly localized metallic state via a quantum metal state and/or a quantum Griffith state.

## Section S12. Point contact spectroscopy

In this section, we describe in detail the results of the point contact spectroscopy performed on device D5. Point contact spectroscopy is a technique commonly used to extract information about the energy gap in a superconductor or also a pseudogap. It involves a measurement of either the contact resistance ($R_{contact,dc}$, defined as $R_{contact} = V_{ac}/I_{ac}$ when $I_{dc} = 0$ A) or the differential conductance ($G_{contact} = I_{ac}/V_{ac}$, see Fig. 4 for measuring geometry) across the contact junction as a function of the dc voltage ($V_{\text{dc}}$) for different temperatures and magnetic fields. The quality of the contact area is of great importance for point contact spectroscopy. Here in this work, the point contact is formed by placing part of the thin film on top of a micron-sized Au electrode. The contact area is determined by the grain size of the sputtered Au electrodes, and it is inevitable that many parallel channels co-exist. Without exerting any additional pressure, the contact is of the van der Waals type (*43–45*). Since multiple channels contribute in parallel, the recorded differential conductance represents a spatial average. The superconducting gap appears as a bump like structure in the differential conductance versus bias voltage when the sample is cooled below the critical temperature. The bump becomes stronger at the temperature is lowered and develops a center dip flanked by the two coherence peaks as seen in panel C and E of Fig. 4 in the main text. The recorded $G_{contact}$ traces typically exhibit such a clear gap signature with a pronounced temperature and magnetic field dependence.

By combining the point contact spectroscopy results with in-plane resistance measurements on the same sample, it is possible to exclude extrinsic or thermal effects as the origin of the feature observed in the point contact spectroscopy. Fig. S16A plots the temperature dependence of the resistance for contact #1 ($R_{contact,dc}$, black line) together with the in-plane sheet resistance ($R_{\text{sheet}}$, blue line). These data were taken simultaneously. The dashed line marks the transition temperature $T_{c,\text{onset}}$ of ~7.5K. The four terminal sheet resistance exhibits the typical superconducting transition and it drops down to zero. The contact resistance keeps increasing before it turns down near 3K. A small kink near 7-8K agrees well with the superconducting transition point in $R_{\text{sheet}}$. Fig. S16B presents the corresponding differential resistance data as a function of the dc bias current $I_{\text{dc}}$: $R_{contact}$ for contact #1 as a black line and the in-plane differential resistance $dV/dI_{\text{sheet}}$ in blue. A single peak is observed in $R_{contact}$ near zero $I_{\text{dc}}$, while $dV/dI_{\text{sheet}}$ remains constant and close to zero within the explored current range of 2μA. The latter demonstrates that the sample remains in the superconducting ground state within this range of bias currents. If we assume the behavior of $R_{contact}$ with bias current in Fig. S16B is the result of a bias current related thermal effect, we would anticipate similar behavior in the temperature dependence of $R_{contact,dc}$ shown in Fig. S16A. However, both curves are qualitatively very different and hence thermal effects can be excluded.

The evolution of the differential conductance $G_{contact}$ of contact #1 with an applied perpendicular magnetic field is shown in Fig. S16C. The magnetic field varies from 0 to 6 T. The temperature dependence can be found in Fig. 4C of the main text. To extract information about the superconducting gap, the extended Blonder–Tinkham–Klapwijk (BTK) model can be used to fit the $G_{contact}$ data (*43–45*, *48*). From the present point contact spectroscopy data, it is difficult to determine the symmetry of the superconducting gap, i.e. whether we are dealing with isotropic *s*-wave, or nodal-like pairing such as anisotropic *s*-wave, *p*-wave or *d*-wave pairing. Previous results obtained on bulk materials provided strong evidence for unconventional $s^{\pm}$-wave pairing (*46*, *47*). We therefore resort to an extended single-band isotropic *s*-wave BTK model to fit the normalized $G_{contact}$ data in Fig. S16D. The normalization of the $G_{contact}$ data proceeds, as usual, by subtracting data obtained at some temperature above the superconducting transition temperature. Here, either data at 8K or 15K were used for this purpose. These data sets are vertically shifted until the data points at higher bias voltage match with those of the low-temperature data. Subsequently, the shifted data sets are used to normalize the low temperature data by division. Examples of normalized data traces are shown in Fig. S16D. The fit procedure yields the following parameters: the gap size $\Delta$, the parameter describing the potential barrier $Z$, and a phenomenological broadening factor $\Gamma$. Fits to the data traces using the BKT model yield the following approximate gap values: 3.0 meV (left panel) and 3.1 meV (right panel) using 15K and 8K data for normalization respectively). This result confirms the validity of the normalization procedure as there is no significant influence of the high temperature data set used. The temperature dependence of the differential conductance $G_{contact}$ of contact #1 plotted in Fig. 4B suggests that the contact is in the insulating regime. The feature associated with a gap persists up to temperatures much higher than $T_{c,\text{onset}}$ and in Fig 4D there is a deviation from the BCS relation. Such behavior is unanticipated and consistent with pseudogap behavior. In contrast, the conductance data for contact #2 in Fig. 4E does not exhibit such pseudogap behavior. For this data set we are in the metallic regime and we observe Andreev reflection and enhanced conductance below the superconducting gap. These two different behaviors in the insulating and metallic regime are not well understood. It possibly reflects some extrinsic effect such as a gap distribution over a wide energy range due to spatial inhomogeneities (*49*). The fitted gap amplitudes are summarized in Fig. 4D and F for both contacts. The phenomenological broadening parameter $\Gamma$ is shown in Fig. S16E for contact #1 and #2. We note that extrinsic effects are "absorbed" in the phenomenological broadening parameter $\Gamma$ in this BTK fitting, so that they have little influence on the gap size. For the sake of completeness, we note that the pseudogap behavior has been reported in a number of previous studies, such as low-density superconductors (*64*) and cuprates (*65*).

Fig. S17 presents point-contact data at different temperatures and magnetic fields both in the insulating regime at a low hole density of $-2.2 \times 10^{12}$ cm$^{-2}$ and in the superconducting regime for a hole density of $-1.07 \times 10^{13}$ cm$^{-2}$. Gap-like behavior is observed even in the insulating regime. This is not anticipated at first. However, as a result of the spatial inhomogeneity superconducting puddles may exist in the contact area even in the insulating states and produce superconducting gap behavior with a reduced gap value. In these two regimes, the above normalization method brings large uncertainty when performing the BTK fittings, as the spectra at higher bias voltage deviate a lot from 1. Nonetheless, rough estimation of the gap size yields the gap values of $2.3 \pm 0.2$ meV and $1.8 \pm 0.2$ meV for insulating and hole-doped superconducting regime respectively, as marked with dashed black lines in Fig. S18. The corresponded gap ratios ($\Delta/k_B T_{c,\text{onset}}$) are 3.5 and 2.8 respectively, hence both exceed the BCS limit, supporting the strong pairing interaction in this hole-doped superconducting regime as well.

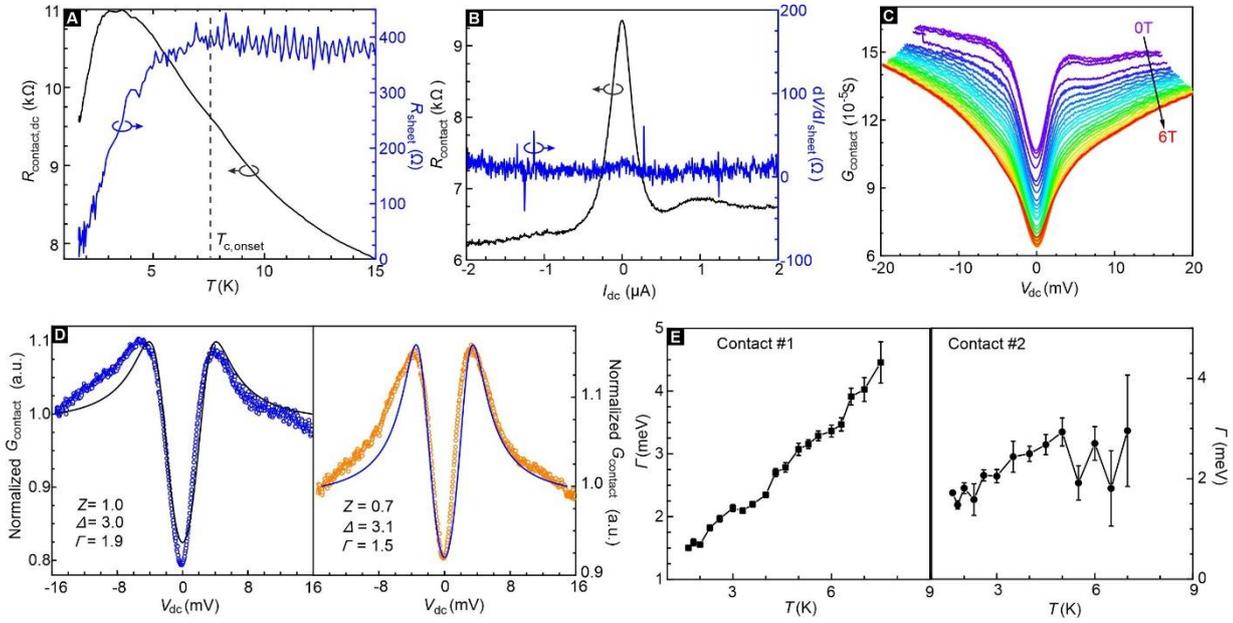

**Fig. S16. Analysis of the point-contact spectroscopy data.** All data shown were recorded on device D5 for an electron density of $1.38 \times 10^{13}$ cm$^{-2}$. (**A**) Temperature dependence of the contact resistance $R_{contact,dc}$ for contact #1 (black line) and the in-plane sheet resistance $R_{sheet}$ (blue line). The dashed line marks the onset of the superconducting transition ($T_{c,\text{onset}}$). (**B**) Differential resistance $R_{contact}$ for contact #1 (black line) and the in-plane differential sheet resistance $dV/dI_{sheet}$ (blue line) as a function of the bias current $I_{dc}$. (**C**) Differential conductance $G_{contact}$ as a function of the dc voltage $V_{dc}$ for different values of the magnetic field from 0T to 6 T in 0.5 T steps. Data are recorded on contact #1 at a fixed temperature of 1.65 K. (**D**) Fit of the BTK model to the differential conductance data recorded at 1.65 K for contact #1. The data is either normalized using 15 K (left panel) or 8 K (right panel) high temperature data. (**E**) Temperature dependence of the broadening parameter $\Gamma$ extracted from the BTK fit. Data are shown for both contact #1 (left panel) and contact #2 (right panel).

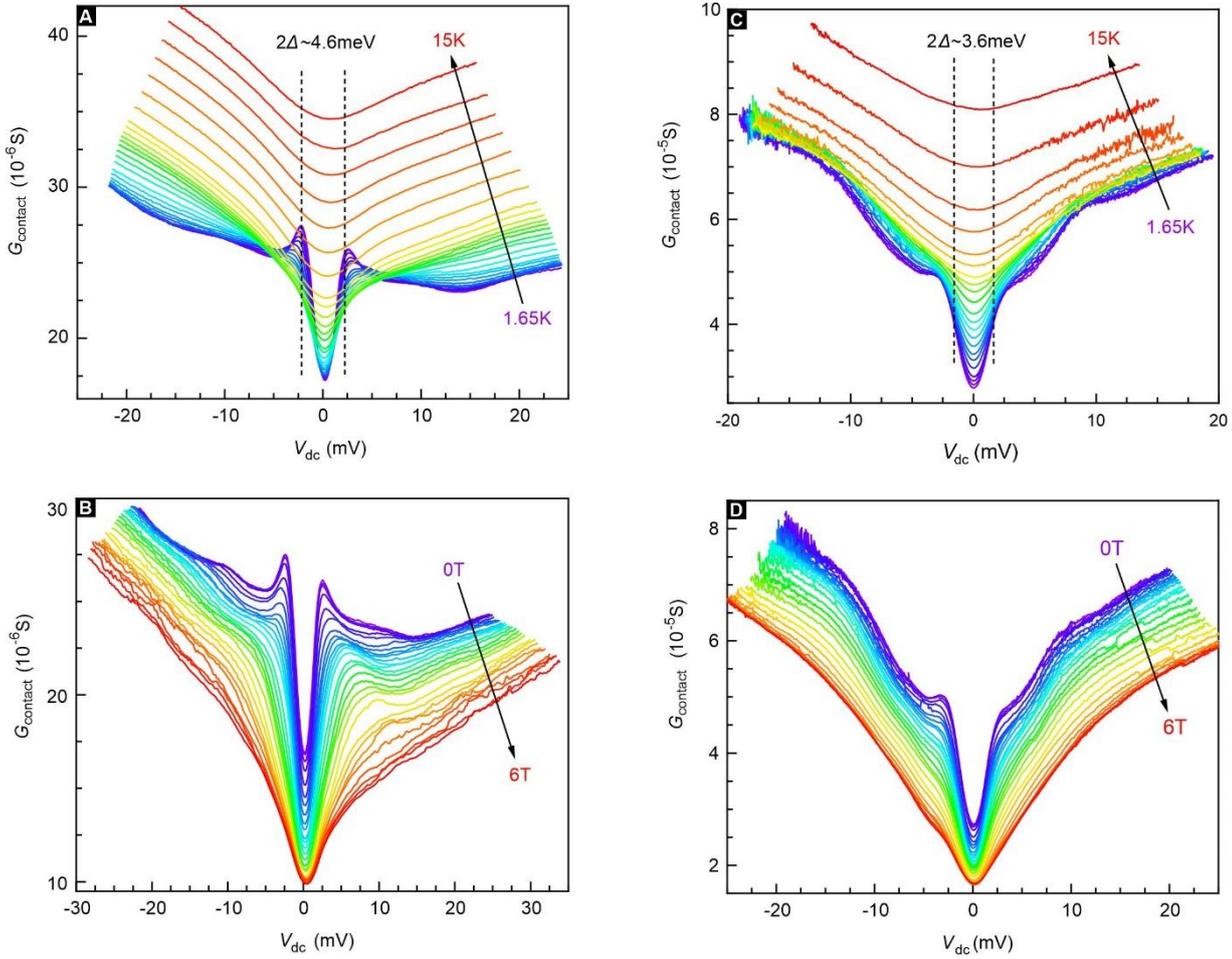

**Fig. S17. Point-contact spectroscopy in the insulating and hole-doped superconducting regime.** Data were recorded on contact #1 of device D5. **(A)** Temperature evolution of the differential conductance $G_{contact}$ recorded as a function of $V_{dc}$ for $B = 0$ T and a low net hole density of $-2.2 \times 10^{12}$ cm$^{-2}$. The temperature is varied from 1.65K to 15K. The dashed lines demarcate the gap region. **(B)** Differential conductance $G_{contact}$ as in panel **A** but for different values of the perpendicular magnetic field (0T to 6T) at a fixed temperature of 1.65 K. **(C and D)** Same as (**A and B**) but for a hole density of $-1.07 \times 10^{13}$ cm$^{-2}$, respectively.

## Section S13. Monolayer device showing no evidence of a bulk gap

In this section, we discuss for the sake of completeness a 1T'-MoTe$_2$ monolayer device showing transport behavior different from all other measured monolayer samples in this work (device D6). It exhibits superconductivity for the full range of gate voltages without any sign for the emergence of a bulk gap. The specific reasons for why this sample behaves distinct from all other monolayer devices addressed in this work are unknown. The disorder induced density in this sample is on the order of 10$^{12}$ cm$^{-2}$. The behavior of this sample resembles what has been reported previously in the literature (*20*). While this sample was fabricated using the same processing steps as all other samples, it was left on the Si substrate surface for approximately 5 days prior to protective encapsulation with a thin hBN flake and subsequent transfer onto a substrate with pre-patterned electrodes. Even though the sample was stored inside the argon atmosphere of the glove box, some sample degradation may have occurred during this long delay between exfoliation and encapsulation. Fig. S18A displays an optical image of the device. In panel B the band structure schematic for a semi-metal is shown that appears to be applicable for this device. Transport results are shown in this figure as well: $R$ versus $T$ (C), $R$ versus $B$ (D) and the differential resistance d$V$/d$I$ as a function of the dc bias current $I_{dc}$ (E). In all cases, the charge carrier density serves as an additional parameter. The thick black trace in Fig. S18C corresponds to average zero carrier density using the criterion that at the applied gate voltages, the resistance exhibits a peak in the normal state at 10 K. At this gate voltage condition the critical current also reaches its minimum (Fig. S18E). Fig. S18F plots the temperature at the transition of superconductivity $T_{c,0}$ as well as the critical current $I_c$ as a function of the charge carrier density.

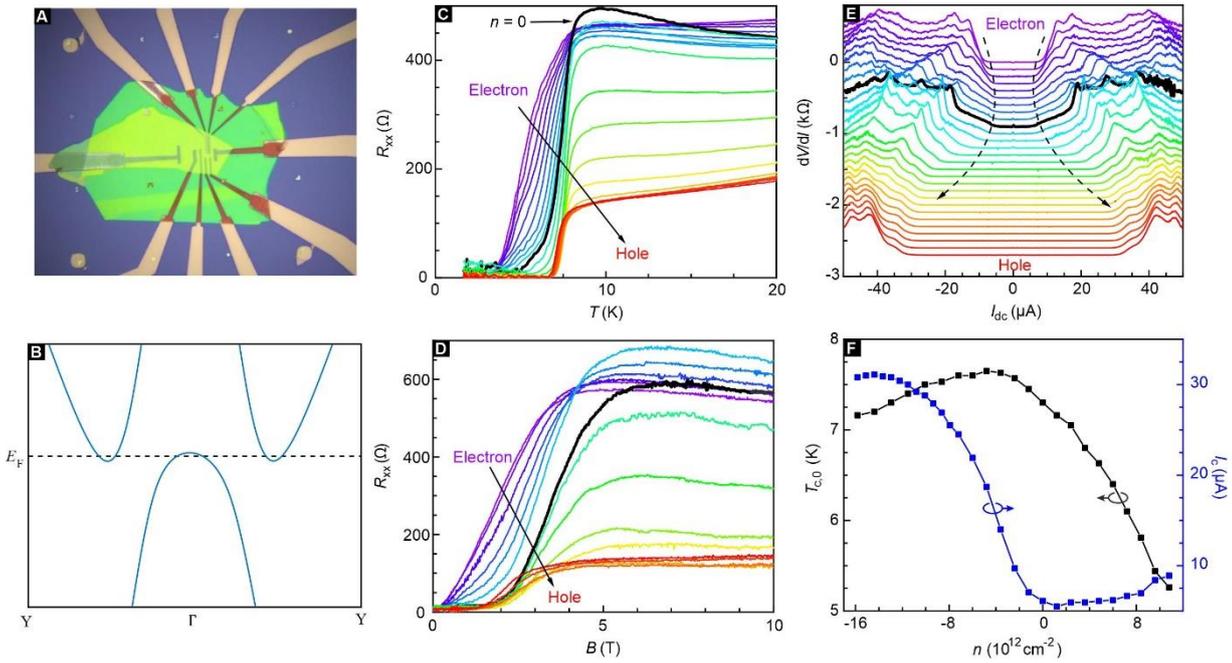

**Fig. S18. Monolayer device showing no evidence of a bulk gap** (device D6). (**A**) Optical image of the device. (**B**) Schematic band structure for a semi-metal as applicable to this device. (**C**) Temperature dependence of the sheet resistance for different carrier densities. Data are recorded in the absence of a magnetic field. The thick black line is recorded at gate voltages corresponding to zero average density. (**D**) Sheet resistance as a function of the perpendicular magnetic field. The charge carrier density serves as an additional parameter. All data taken at 1.65 K. (**E**) Differential resistance as a function of the dc-bias current. The charge carrier density serves as an additional parameter. These data are used to extract the critical current $I_c$ in panel **F**. Data are recorded at $T = 1.65$ K and $B = 0$ T. (**F**) Temperature at the transition of superconductivity, $T_{c,0}$, and critical current, $I_c$, as a function of the average carrier density. The density range in panel **C-F** is $-1.51 \times 10^{13} < n < 1.08 \times 10^{13}$ cm$^{-2}$.

# Section S14. Comparison with other 2D superconductors

Key parameters of our superconducting samples are compared with other 2D superconductors in the table below.

| 2D SC | $T_{c,onset}$ (K) | $n$ (cm$^{-2}$) | Gate range (cm$^{-2}$) | $B_{c2,\perp}$ (T) | $B_{c2,\parallel}$ (T) | Hole or electron SC | Inversion symmetry | Out of plane symmetry |
|---|---|---|---|---|---|---|---|---|
| (type-I) 2H-NbSe$_2$(35) | ~3.2 | >10$^{14}$ | Not applicable | ~1 | >30 (~6$B_p$) | hole | no | yes |
| (type-I) 2H-TaS$_2$(26) | ~2.2 | >10$^{14}$ | Not applicable | ~1 | >30 (~6$B_p$) | Not known | no | yes |
| (type-I) 2H-MoS$_2$(1) | ~10 | ~7×10$^{13}$ | liquid gate ~10$^{14}$ | 0.5 | >50 (~6$B_p$) | electron | no | yes |
| (type-II) Stanene(52) | ~1.5 | >10$^{14}$ | Not applicable | ~0.6 | ~4 (~4$B_p$) | two hole bands | yes | no |
| (type-II) PdTe$_2$(53) | ~0.7 | >10$^{14}$ | Not applicable | ~0.8 | ~20 (~7$B_p$) | two bands | yes | no |
| TBLG(8) | ~2 | ~1.4×10$^{12}$ | 2×10$^{13}$ | ~0.1 | ~1 (~1$B_p$) | ambipolar | yes | no |
| TTLG(7) | ~3 | ~3×10$^{12}$ | 2×10$^{13}$ | ~0.4 | ~10 (~3$B_p$) | ambipolar | no | yes |
| 1T'-WTe$_2$ (9, 10) | ~1.2 | ~5×10$^{12}$ | 2×10$^{13}$ | ~0.05 | ~4.5 (~6$B_p$) | electron | yes | no |
| **1T'-MoTe$_2$** | **~7.5** | **~5×10$^{12}$** | **2×10$^{13}$** | **~2** | **~20 (~2$B_p$)** | **ambipolar** | **yes** | **no** |

**Table S1. Comparison of key parameters of the superconducting** 1T'-MoTe$_2$ **monolayers with some other 2D superconductors, that have been reported in the literature.** They include type-I and type-II Ising superconductors, twisted graphene layers (TBLG = twisted bilayer graphene, TTLG = twisted trilayer graphene) and monolayer 1T'-WTe$_2$. In some cases the density is not known.